\documentclass[aps,prd,showpacs,notitlepage,nofootinbib,superscriptaddress,floatfix,showkeys,twocolumn]{revtex4-1}
\usepackage{dcolumn}
\usepackage{blindtext}
\usepackage{hyperref}
\usepackage{amsmath,amssymb}
\usepackage{float}
\usepackage{microtype}
\usepackage{graphicx}
\usepackage{capt-of}
\usepackage{bm}
\usepackage{latexsym}
\usepackage{epsfig}
\usepackage{psfrag}
\usepackage{color}
\usepackage[dvipsnames]{xcolor}
\usepackage{subfigure}
\usepackage{subcaption}
\usepackage[section]{placeins}
\usepackage{natbib}
\usepackage{ulem}
\usepackage{cancel}
\usepackage{lipsum}
\usepackage{comment}
\usepackage{booktabs}
\usepackage{csquotes}

\def\red{\textcolor{red}}
\newcommand{\arp}[1]{\red{{[AR: #1]}}}

\def\GeV{\mathrm{GeV}}
\def\mH{\mathcal{H}}
\def\HI{H_{\rm end}}
\def\Mp{\mathrm{M}_{\rm Pl}}
\def\Tre{T_{\rm reh}}

\def\gst{g_{\rm re}^{*}}
\newcommand{\dd}{\text{d}}

\def\ae{a_{\rm end}}
\def\Ora{\Omega_{\rm rad}}

\def\nn{\nonumber}
\def\l{\left}
\def\r{\right}
\def\vx{\mathbf{x}}
\newcommand{\Eq}[1]{Eq.~(\ref{#1})}
\newcommand{\Eqq}[1]{Eqs.~(\ref{#1})}

\newcommand{\Fig}[1]{Fig.~\ref{#1}}

\def\wphi{\omega_{\phi}}
\def\Gp{\Gamma_\phi}
\def\tGp{\bar{\Gamma}_\phi}
\def\rhop{\rho_\phi}
\def\rhor{\rho_{\rm rad}}

\def\trhop{\tilde{\rho}_\phi}
\def\trhor{\tilde{\rho}_{\rm rad}}
\def\tE{\tilde{\bf{E}}}
\def\tB{\tilde{\bf{B}}}
\def\tJ{\tilde{\bf{J}}}
\def\vB{\mathbf{B}}

\def\vE{\mathbf{E}}
\def\vJ{\mathbf{J}}
\def\vur{\mathbf{u}_{\rm r}}
\def\vup{\mathbf{u}_{\phi}}

\def\teta{\bar{\eta}}
\def\etae{\eta_{\rm end}}

\def\rk{r_{k_*}}
\def\As{A_{\rm s}}

\def\ere{\eta_{\rm reh}}

\def\d{\mathrm{d}}

\def\rhoc{\rho_{\rm cri}}

\def\hij{h_{ij}}
\def\bhij{\bar{h}_{ij}}

\def\bTij{\bar{T}_{ij}}
\def\rhogw{\rho_{\rm GW}}
\def\ogw{\Omega_{\rm GW}}

\def\gseq{g^{*}_{\rm eq}}
\def\gsp{g^{*}_0}
\def\gsre{g^{*}_{\rm re}}
\def\Ophi{\Omega_\phi}
\def\ke{k_{\rm end}}

\def\Hz{\rm Hz}
\def\Mpc{\mathrm{Mpc}}

\begin{document}
\title{Magnetohydrodynamics During Reheating}
\author{Sovan Sau}
\email{E-mail: sovan.sau@gmail.com}
\affiliation{Department of Physics, Indian Institute of Technology, Guwahati, 
Assam, India}
\author{Subhasis Maiti}
\email{E-mail: subhashish@iitg.ac.in}
\affiliation{Department of Physics, Indian Institute of Technology, Guwahati, 
Assam, India}
\author{Sahel Dey}
\email{E-mail: saheldey917@gmail.com}
\affiliation{School of Science, University of Newcastle, University Drive, Callaghan, NSW 2308, Australia}
\author{Alberto Roper Pol}
\email{E-mail: alberto.roperpol@unige.ch}
\affiliation{D\'epartement de Physique Th\'eorique, Universit\'e de Gen\`eve, CH-1211 Gen\`eve, Switzerland}
\author{Debaprasad Maity}
\email{E-mail: debu@iitg.ac.in}
\affiliation{Department of Physics, Indian Institute of Technology, Guwahati, 
Assam, India}

\begin{abstract}
Reheating is a crucial phase that connects inflation with the hot big-bang Universe. During this phase, the nonlinear interactions among the inflaton, the radiation plasma, and the primordial magnetic fields can influence the subsequent evolution of cosmic structures we observe today. In this paper, we develop a multifluid framework to investigate the evolution of three-dimensional nonlinear magnetohydrodynamics (MHD) during the non-instantaneous reheating phase driven by the perturbative decay of the inflaton into the radiation plasma. For the present paper, we consider the simplest scenario where the inflaton field homogeneously injects energy into the radiation sector. We study the MHD evolution of the coupled inflaton–radiation system during the reheating phase, including the evolution of their individual kinetic energies, the velocity perturbations, vorticity, magnetic fields, and gravitational waves. We find that homogeneous energy transfer from the inflaton to the radiation sector increases the radiation and kinetic energy densities but does not significantly amplify the initial magnetic field perturbations. Consequently, the magnetic spectrum remains largely unchanged with respect to the instantaneous reheating scenario, while the velocity perturbations undergo a scale redistribution that affects the resulting GW spectrum. Our results demonstrate the importance of nonlinear multifluid MHD dynamics in understanding the evolution of primordial fields during reheating and provide a framework for exploring more realistic scenarios involving inhomogeneous decay processes and detailed thermalization effects.
\end{abstract}

\maketitle

%=====================================================
\noindent
\section{Introduction}
Reheating is a crucial phase in the early Universe when the inflaton field gradually decays into radiation,
leading to a radiation-dominated Universe. During inflation, the energy density of the Universe is dominated by the inflaton field. After inflation ends, the inflaton starts to oscillate around its potential minimum and gradually transfers its energy into relativistic particles.
This energy transfer leads to the formation of a thermal bath and sets the initial conditions for the subsequent evolution of the Universe. Therefore, understanding the dynamics of reheating is important not only for the thermal history of the Universe but also for understanding the properties of the primordial plasma.

Most studies of reheating focus on the evolution of the homogeneous energy densities of the inflaton and radiation fields or use linear perturbation theory~\cite{Allahverdi:2010xz, RevModPhys.78.537, Amin:2014eta, Hamazaki:1996ir, Kodama:1984ziu}. However, the thermalization process is a complex nonlinear phenomenon involving energy transfer, fluid motion, and the development of small-scale structures. How the radiation plasma becomes thermalized, how velocity perturbations evolve, and how different components of the plasma interact during this period are some of the important questions still not fully understood. A detailed study of these processes requires going beyond the homogeneous approximation and considering the nonlinear evolution of the coupled system.

Magnetic fields provide an important probe of the early-Universe plasma. They are observed on different astrophysical scales, including galaxies, galaxy clusters, filaments of the large-scale structure~\cite{Carretti:2022fqk}, and possibly cosmic voids~\cite{Neronov:2010gir, MAGIC:2022piy}. However, their origin remains an open problem~\cite{Durrer:2013pga, Subramanian:2015lua, Vachaspati:2020blt}. Primordial magnetic fields generated during inflation provide one possible explanation~\cite{Widrow:2002ud, Durrer:2013pga}. Their evolution after inflation requires determining their non-linear dynamics. Once the radiation plasma is formed during reheating, its high conductivity couples the magnetic field with the fluid motion, making magnetohydrodynamics (MHD) the natural framework to study their evolution \cite{Brandenburg:1996fc, Ahonen:1996nq, Arnold:2000dr, RoperPol:2025lgc, Iarygina:2026bom}.

In a conducting plasma, magnetic fields and fluid motion evolve together through nonlinear interactions. Processes such as advection, stretching, and energy transfer between magnetic and kinetic components can modify the evolution of both sectors. Previous studies have suggested that dynamo mechanisms during reheating may amplify primordial magnetic fields~\cite{Calzetta:1997ku, Calzetta:2010zu, Calzetta:2015hda, Durrer:2013pga, Iarygina:2026bom}.
%\arp{comment about new paper by Oksana, Axel} 
However, the efficiency of dynamo amplification strongly depends on the presence and properties of velocity perturbations and turbulence. For a spatially homogeneous inflaton decay, the decay process does not directly generate spatial variations in the radiation plasma. Therefore, in the homogeneous reheating scenario, inflaton energy does not source inhomogeneous velocity perturbations, and the direct driving of the dynamo amplification is expected to be limited. Nevertheless, continuous injection of radiation during reheating can strongly affect the plasma dynamics by changing the evolution of kinetic energy, vorticity, and small-scale turbulent motion.
% \red{When the inflaton is inhomogeneous, for example, in axion inflation with strong backreaction,
% dynamo amplification of the magnetic field has been recently reported \cite{Iarygina:2026bom}.}

The nonlinear evolution of the fluid small-scale structures is important because it controls the transfer of energy across different length scales and determines how the primordial plasma evolves towards thermal equilibrium. Such effects cannot be captured by linear perturbation theory or by considering only the background evolution. A complete MHD treatment is required to understand the coupled evolution of the radiation fluid and electromagnetic fields during reheating.

In this work, we study the nonlinear MHD evolution of the inflaton–radiation system during reheating using a multi-fluid framework. We numerically evolve the coupled inflaton field, radiation plasma, and electromagnetic fields in an expanding Universe using the open-source {\sc Pencil Code} \cite{PencilCode:2020eyn}, which we have adapted to include the inflaton–radiation system. As a minimal setup, we consider homogeneous inflaton decay, allowing us to study the effect of continuous energy injection without introducing additional velocity perturbations from inhomogeneous decay. We find that, in this scenario, reheating has a negligible effect on the magnetic field, while it leads to non-trivial evolution of the kinetic energy, vorticity, and turbulent plasma motions. This provides a first step towards a more complete MHD description of reheating and its impact on the evolution of fluid and magnetic fields in the early Universe. The framework developed here can be extended to include more realistic effects, such as inhomogeneous inflaton decay, detailed particle interactions, and thermalisation processes. Such extensions will help us better understand the transition from inflaton domination to the hot plasma state and the evolution of primordial fields, dark matter, radiation, and small-scale structures during this important epoch.
%=====================================================

\section{MHD during reheating}

We model reheating as a two-component system consisting of a neutral inflaton field and a radiation fluid. For simplicity, the inflaton is treated as a condensed, spatially homogeneous field, $\phi=\phi(t)$. The radiation fluid is composed of SM particles produced through the decay of the homogeneous inflaton $\phi$. All the SM species are assumed to be strongly interacting, described as a single effective fluid. In this setup, primordial magnetic fields originate from quantum fluctuations during inflation with an initially tiny seed energy density. As the inflaton field is assumed to be electrically neutral and no additional couplings to the gauge sector are introduced, the magnetic field couples naturally only to the radiation fluid.

\subsection{Equations of motion}

The total energy-momentum tensor satisfies $\nabla_\alpha T^{\alpha\beta}=0$.
Allowing for energy exchange between the
inflaton $\phi$
and the radiation (rad) fluid, we write
\begin{equation} \label{eom_tot}
    \nabla_{\alpha}T^{\alpha\beta}_{\rm rad} = - 
    \nabla_\alpha T^{\alpha\beta}_{\phi}
    = -(1+\wphi)
    \Gamma_{\phi}\, T^{\alpha \beta}_\phi u_\alpha^\phi ,
\end{equation}
with $T^{\alpha\beta}=(p+\rho)u^\alpha u^\beta + pg^{\alpha\beta} +\pi^{\alpha\beta}$ denoting the energy-momentum tensor of a dissipative fluid, where $\pi^{\alpha\beta}$ is the traceless shear-stress tensor describing dissipative, anisotropic viscous stresses. It satisfies $u_\alpha\pi^{\alpha\beta}=0$ and $g_{\alpha\beta}\pi^{\alpha\beta}=0$~\cite{Landau1987Fluid, Romatschke:2009im, Andersson_2007}.
%where $\pi^{\alpha\beta}$ is the traceless shear-stress tensor describing dissipative, anisotropic viscous stresses. In the Landau frame, it is given by $\pi^{\alpha\beta} =\Delta^{\alpha\beta}_{\mu\nu}T^{\mu\nu}$, with $\Delta^{\alpha\beta}_{\mu\nu} =\frac{1}{2}(\Delta^\alpha_\mu\Delta^\beta_\nu +\Delta^\alpha_\nu\Delta^\beta_\mu) -\frac{1}{3}\Delta^{\alpha\beta}\Delta_{\mu\nu}$, $\Delta_{\mu\nu}=g_{\mu\nu}+u_\mu u_\nu$, and satisfies $u_\alpha\pi^{\alpha\beta}=0$ and $g_{\alpha\beta}\pi^{\alpha\beta}=0$~\cite{Landau1987Fluid, Romatschke:2009im, Andersson_2007}.
Here, $u^\alpha_\phi$ is the four-velocity of the inflaton fluid satisfying the normalisation condition $u^\alpha_\phi u^\phi_\alpha=-1$. The isotropic and homogeneous expanding Universe is described by the Friedmann-Lema\^itre-Robertson-Walker (FLRW) metric tensor, $g_{\alpha\beta} = a^2(\eta)\, \text{diag}(-1,1,1,1)$. For a homogeneous inflaton energy density, the inflaton velocity is at rest, $u^\mu_\phi=(a^{-1},0,0,0)$. The radiation four-velocity is $u^\mu_{\rm rad} = a^{-1} \gamma (1,\vur)$, with $\gamma^2 = 1/(1 - \vur^2)$, also satisfying $u_{\rm rad}^\alpha u_\alpha^{\rm rad} = -1$. The parameter $\Gamma_\phi$ characterises the inflaton decay rate. The inflaton and radiation equations of state (EoS) are 
\begin{equation}
    \wphi \equiv \frac{p_\phi}{\rho_\phi}= \frac{\dot{\phi}^2-V}{\dot{\phi}^2+V}, \qquad
    \omega_{\rm rad}\equiv \frac{p_{\rm rad}}{\rho_{\rm rad}}=\frac{1}{3}.
\end{equation}

To study the evolution of a radiation-dominated plasma, it is convenient to work in comoving coordinates and conformal time $\eta$~\cite{Brandenburg:1996fc}. We define the comoving energy densities $\trhop$ and $\trhor$, electromagnetic fields $\tE$ and $\tB$, and current density $\tJ$ as
\begin{subequations}
\begin{align}
    \trhop = a^{3(1+\wphi)}\rhop,\quad 
    \trhor = a^4 \rhor,
    \\
    \tE=a^2 \vE,\quad 
    \tB= a^2 \vB, \quad 
    \tJ= a^3 \vJ.
\end{align}
\end{subequations}

Combining Maxwell equations with energy-momentum conservation [cf.~\Eq{eom_tot}], we obtain the coupled dynamical equations in the limit of subrelativistic radiation velocity $\vur^2 \ll 1$ \cite{RoperPol:2025lgc},
\begin{subequations}\label{eq:fluid}
\begin{align}
    \frac{\partial \ln \trhop}{\partial \eta} 
    &\, =  
    - \mathcal{S}_\phi(\eta),\label{eq:3a}
    \\
    \frac{\partial \ln \trhor}{\partial \eta} 
    &\, =  - \frac{4}{3}\left( \nabla\cdot \vur + \tfrac{1}{2}
    \vur\cdot\nabla \ln \trhor \right) 
    \nonumber \\ &\, + \frac{1}{\trhor}\left[\tilde \eta_{\rm M} \tJ^2
    - \vur \cdot (\tJ \times \tB)\right]  
    + \mathcal{S}_r(\eta), \label{eq:3b}
    \\
    \frac{\partial \vur}{\partial \eta} = &\, - \vur \cdot \nabla \vur +
    \frac{\vur}{3} ( \nabla\cdot \vur + \tfrac{1}{2} \vur \cdot \nabla \ln \trhor ) \nonumber \\ &\, - \frac{\vur}{\trhor} \left[\tilde \eta_{\rm M} {\tJ}^2 + \tfrac{1}{2} \vur (\tJ \times \tB)\right] -\frac{1}{4} \nabla
    \ln \trhor \nonumber \\ &\, + \frac{3}{4 \trhor} (\tJ\times \tB) + \frac{2}{\trhor} \nabla \cdot (\trhor\,
    \tilde{\nu}\, {\pmb S}), \label{eq:3d} \\
    \label{eq:induction}
    \frac{\partial\tB}{\partial \eta}
    = &\, \nabla\times(\vur\times \tB)
    +
    \tilde \eta_{\rm M}
    \nabla^2\tB, \quad \tJ = \nabla \times \tB,
    \end{align}
\end{subequations}
where $\mathcal{S}_\phi=a(1+\wphi) \Gp$ and 
$\mathcal{S}_{\rm r}=a^{1-3\wphi}\frac{\trhop}{\trhor}\mathcal{S}_\phi$~\cite{Giudice:2000ex, Erickcek:2011us, Erickcek:2015jza} model the interactions between the radiation and inflaton field, $\tilde \eta_{\rm M} = a^{-1} \eta_{\rm M}$ and $\tilde{\nu} = a^{-1} \nu$ are the comoving magnetic diffusivity and shear viscosity, and $S_{ij} = \tfrac{1}{2} (u_{i,j}^{\rm r} + u_{j,i}^{\rm r}) - \tfrac{1}{3} u_{l,l}^{\rm r} \delta_{ij}$ is the traceless rate-of-strain tensor. 

The evolution of the scale factor is determined by the Friedmann equation,
\begin{equation}
    \mH =
    \frac{a^{-1}}{\sqrt{3}\Mp}\l( a^{1-3\wphi} \left< \trhop \right>
    + \left< \trhor \right>+\langle\tilde{\rho}_{\rm EM}\rangle \r)^{1/2},\label{eq:scale}
\end{equation}
where ${\cal H} = \partial_\eta \ln a$ is the conformal Hubble rate, related to the physical Hubble rate $H$ as ${\cal H} = a H$, angular brackets indicate an average over the simulation volume, and $\Mp \simeq 2.43 \times 10^{18}$ GeV is the reduced Planck mass. Here $\tilde{\rho}_{\rm EM}$ is the comoving electromagnetic energy density.

\subsection{Evolution of the background energy density}

At the onset of reheating, corresponding to $\eta/a \sim H^{-1} \ll \Gamma_\phi^{-1}$, the Universe is predominantly inflation-dominated. Neglecting inhomogenous
contributions, the evolution equations for the comoving background energy densities reduce to
\begin{subequations}
\label{eq:back_evo}
\begin{align}
    \frac{\partial \trhop}{\partial \eta}
    &= -a(1+\wphi)\,\trhop\,\Gp, \\%[4pt]
    \frac{\partial \trhor}{\partial \eta}
    &= a^{\,2-3\wphi}(1+\wphi)\,\trhop\,\Gp.
\end{align}
\end{subequations}

To obtain analytical solutions, we assume that the effective equation of state of the inflaton remains constant throughout reheating. Under these assumptions, the comoving inflaton energy density evolves as
\begin{align}\label{eq:phi_evo}
    \frac{\trhop (\bar \eta > 1)}{\tilde{\rho}_{\phi}^{\rm end}}\simeq \exp\l[ \frac{1+3\wphi}{3(1+\wphi)}\tGp\l(1-{\teta}^{\frac{3(1+\wphi)}{1+3\wphi}}\r) \r],
\end{align} 
where $\trhop^{\rm end}=3\,\ae^{3(1+\wphi)}\HI^2\Mp^2$ denotes the inflaton energy density at the end of inflation, and $\tGp=\Gp/\HI$ is the normalized decay rate of the inflaton field. To solve the above equation, we introduce the dimensionless conformal time $\teta=\eta/\etae$. In the limit $\tGp\ll1$, we find that the comoving energy density of the inflaton field initially remains approximately constant.
This behaviour persists as long as $\tGp{\teta}^{\frac{3(1+\wphi)}{1+3\wphi}}\ll 1$.
During this phase, the inflaton dominates the energy of the Universe, and the
scale factor evolves as $a/\ae = \bar \eta^{2/(1 + 3 \wphi)}$.

The corresponding evolution of the comoving radiation energy density is given by
\begin{align}
    \frac{\trhor (\bar \eta > 1)}{\tilde{\rho}_{\phi}^{\rm end}}
    \simeq
    \Ora^{\rm end}+\frac{1+3\wphi}{5-3\wphi}
    \, \tGp \l(\teta^{\frac{5-3\wphi}{1+3\wphi}}-1\r),
\end{align}
where $\Ora^{\rm end} \equiv
\rhor^{\rm end}/(\rho_\phi^{\rm end} + \rhor^{\rm end})
\simeq \rhor^{\rm end}/\rho_\phi^{\rm end}$
describes the initial radiation energy density at the end of inflation.
This expression explicitly demonstrates that the comoving
radiation
bath grows continuously during reheating rather than being produced instantaneously.
For a matter-dominated reheating phase ($\wphi=0$), the radiation energy density evolves as $\trhor\propto a^{5/2}$, whereas for a radiation-like expansion ($\wphi=1/3$),
it scales as $\trhor\propto a^{2}$.
Such a gradual buildup of the thermal bath can significantly influence the evolution of small-scale dynamics of the radiation plasma during reheating.

Throughout this work, we assume both the inflaton decay rate and the effective equation of state to be constant. In a more realistic reheating scenario, however, the background temperature evolves continuously, while the effective inflaton mass may also vary with time. Consequently, both the decay rate and the effective equation of state are, in general, time dependent. Nevertheless, the constant-parameter approximation adopted here provides a useful first step towards implementing a multi-fluid reheating system in lattice simulations. Such a framework enables the study of the nonlinear evolution of both the inflaton and radiation fields, together with their associated perturbations. Incorporating time-dependent decay rates and equations of state will be the subject of future work aimed at studying more realistic reheating dynamics.

%Nevertheless, the constant-parameter approximation adopted here provides a useful first step towards implementing a multi-fluid system with nonlinear evolution of both the inflaton and radiation fields, together with their associated perturbations, in lattice simulations. Incorporating time-dependent decay rates and equations of state will be the subject of future work to study more realistic reheating dynamics.

The temperature of the thermal radiation plasma is related
to its energy density as $\rhor=\frac{\pi}{30}\gst T^4$, where $\gst$ denotes the effective number of relativistic degrees of freedom.
As radiation is continuously produced, its energy density
eventually catches up with that of the inflaton,
satisfying $\rhop(\ere)=\rhor(\ere)$ at the reheating temperature.
For reheating scenarios with $\wphi\leq1/3$, this equality is typically reached when $H\simeq \Gp$, whereas for $\wphi>1/3$ it may occur even when $H>\Gp$~\cite{Haque:2022kez}.
For the post-inflationary dynamics considered in this work, with $\wphi=0$,
the reheating temperature follows from \Eq{eq:scale},
setting $\Gamma_\phi = {\cal H}/a$
and $\rho_\phi = \rho_{\rm rad}$,
\begin{align}
\Tre=\l[\frac{15 \,\rhor(\ere)}{\pi \gst}\r]^{1/4}
\simeq \l(\frac{45}{\pi\gst}\r)^{1/4}\sqrt{\Gp\Mp}.
\end{align}

Within canonical slow-roll inflation, the Hubble parameter at the end of inflation can be expressed as $\HI \simeq \pi \Mp \sqrt{\frac{\rk \As}{2}}$,
where $\rk$ and $\As$ denote the tensor-to-scalar ratio and the amplitude of scalar perturbations, respectively. Using the latest \textit{Planck} 2018 constraint, $\rk \lesssim 0.036$, together with $\As\simeq2.1\times10^{-9}$~\cite{Planck:2018vyg}, we obtain $\HI\lesssim 4.81\times10^{13}\,\GeV \simeq 1.9\times10^{-5}\Mp$.
This corresponds to an initial critical energy density $\rhop(\etae)\simeq\rhoc(\etae)=3\HI^2\Mp^2\lesssim4.12\times10^{64}\,\GeV^4$,
where $\etae={\cal H}_{\rm end}^{-1}$ denotes the end of inflation.

Finally, adopting $\gst=106.75$, appropriate when all Standard Model particles are relativistic~\cite{Kolb:1990vq},
and taking a representative decay parameter $\bar \Gamma_\phi \equiv \Gp/H_{\rm end} = 10^{-5}$,
we obtain the corresponding reheating temperature,
\begin{equation}
    \Tre \simeq 10^{12}\,\GeV \sqrt{\tfrac{\bar \Gamma_\phi}{10^{-5}}}.
\end{equation}
In our computation, we assume two sample values of ${\bar \Gamma_\phi} = (10^{-5}, 10^{-7})$ throughout, and the associated reheating temperature would be $\sim (10^{12}, 10^{11})$ GeV.

\subsection{Gravitational waves from fluid and electromagnetic sources}
In this subsection, we discuss the gravitational waves (GWs) generated during reheating by fluid motion and the electromagnetic fields. In the presence of an active source with an anisotropic stress-energy component, the line element, including the tensor perturbation $\hij$, can be written as
\begin{align}
    \dd s^2=a^2(\eta)\l[-\dd \eta^2+(\delta_{ij}+\hij)\dd x^i \dd x^j\r].
\end{align}
Here, $\hij$ denotes the transverse $(\partial_i\hij=0)$ and traceless $(h_{ii}=0)$ component of the metric perturbation. Linearising the Einstein equation and working in Fourier space, the tensor perturbation satisfies~\cite{Grishchuk:1974ny, Caprini:2018mtu, RoperPol:2018sap}
\begin{align}
    \l(\frac{\partial^2}{\partial\eta^2}+k^2-\frac{a''}{a}\r)\bhij
    =
    \frac{2}{\Mp^2}\frac{\bTij^{\rm TT}}{a},
\end{align}
where $\bar{h}_{ij}=a\hij$ and $\bar{T}_{ij}^{\rm TT}=a^2T_{ij}^{\rm TT}$ are the rescaled tensor perturbation and energy-momentum tensor, respectively. The source on the right-hand side corresponds to the transverse-traceless (TT) component of the anisotropic stress tensor.

The expansion of the Universe  modifies, in general,
the dispersion relation $\omega^2 = k^2 - a''/a$.
During the
radiation-dominated era, $a(\eta)\propto\eta$, and hence $a''=0$, so that the %\sout{homogeneous}
tensor modes freely propagate once the source is no longer active.
During reheating, however, the expansion history can be nontrivial. In general, $a''/a$ is
determined by $\wphi$ and the average inflaton energy density ($\langle\rho_\phi \rangle$),
\begin{equation}
    \frac{a''}{a} = \frac{a^2}{6 \Mp^2} \langle \rho_\phi \rangle (1 - 3 \wphi)\,,
\end{equation}
modifying the propagation of the tensor modes through the time-dependent term $a''/a$~\cite{Haque:2021dha}. In the scenario considered here, the inflaton transfers its energy to the radiation sector and generates velocity perturbations in the fluid. Consequently, GWs can be sourced through two distinct channels: the velocity perturbations of the fluid~\cite{Deryagin:1986qq,Hogan:1986qda,Kamionkowski:1993fg,Mack:2001gc,Nicolis:2003tg,Caprini:2006jb,Gogoberidze:2007an,Hindmarsh:2013xza,Hindmarsh:2015qta,Hindmarsh:2017gnf,RoperPol:2018sap,Niksa:2018ofa,
RoperPol:2019wvy,Kahniashvili:2020jgm,Neronov:2020qrl,Jinno:2022mie,Jinno:2022mie,Auclair:2022jod,Dahl:2024eup,RoperPol:2023dzg,Caprini:2024gyk,Caprini:2026nnk}
and the anisotropic stress of the electromagnetic field~\cite{Caprini:2001nb,Caprini:2003vc,Sorbo:2011rz,Caprini:2014mja,Ito:2016fqp,Sharma:2019jtb,Okano:2020uyr,RoperPol:2021xnd,Brandenburg:2021bvg,RoperPol:2022iel,RoperPol:2023bqa,Maiti:2025rkn,Maiti:2025awl,Maiti:2025cbi,Maiti:2025ijr,Ragavendra:2026fgs,Maiti:2026lvx,Maiti:2026hsn, Maiti:2026dsl}.

In principle, both the velocity perturbations of the inflaton field and those of the
radiation field can contribute to GW production, together with the magnetic field.
In our setup, we 
consider the inflaton field as a homogeneous condensate,
hence neglecting its velocity perturbations due to inhomogeneities.
Therefore, in the present case, only the velocity perturbations of the radiation field contribute to GW production, along with the anisotropic stresses associated with the magnetic field. The rescaled energy-momentum tensor can then be written as \cite{Caprini:2006jb,Brandenburg:2017neh,RoperPol:2019wvy,RoperPol:2025lgc,Figueroa:2026esg}
\begin{align}
    \bTij=\frac{4}{3}\trhor \gamma^2 u_i^{\rm r} u_j^{\rm r}
    -\tilde B_i \tilde B_j+\l(\frac{\trhor}{3}+\frac{\tilde 
    B^2}{2}\r)\delta_{ij}\,.
\end{align}
Only the TT component of $T_{ij}$, denoted by 
$T_{ij}^{\rm TT}$, acts as a source for GWs.
Thus, in our computation,
we omit the contributions proportional to $\delta_{ij}$ since they
vanish after applying the TT projection to the energy-momentum
tensor.

The GW energy density, expressed in terms of the rescaled tensor
perturbations $\bar h_{ij}$, is given by \cite{RoperPol:2018sap}
\begin{align}
    \rhogw &=
    \frac{\Mp^2}{4 a^2}
    \langle \hij'\hij'\rangle\nn\\
    &=
    \frac{\Mp^2}{4 a^4}
    \l(\langle \bhij'\bhij'\rangle
    -2\mH\langle \bhij\bhij'\rangle
    +\mH^2\langle\bhij\bhij\rangle\r).
\end{align}
The spectral energy density of the GWs is defined as,
\begin{align}
    \ogw(\eta)=\frac{\rhogw(\eta)}{\rhoc(\eta)}=\int_{-\infty}^{\infty}\ogw(k,\eta)\d\ln k
\end{align}
where $\rhoc (\eta) =3H^2 (\eta) \Mp^2$ is the critical energy density at a time $\eta$ and
$\ogw(k,\eta) \equiv \dd \ogw(\eta)/\dd \ln k$ is the normalized GW energy density spectrum.

After reheating, the Universe enters the standard radiation-dominated era, during which the
GW energy density decays
as $a^{-4}$, as does the radiation energy density.
Accounting for the change in the effective number of relativistic degrees of freedom
and assuming adiabatic expansion,
the present-day GW  spectrum is \cite{Caprini:2018mtu}
\begin{align}
 \ogw(k,&\eta_0)\, h^2= \nonumber \\
    & \l(\frac{\gseq}{\gsre}\r)^{4/3} 
    \l(\frac{\gsp}{\gseq}\r)^{1/3}
    \Omega_{\rm R}\, h^2\, \ogw(k,\ere),
\end{align}
where $h$ is the present-day Hubble rate divided by 100 km/s/Mpc.
Here, $\gsre\simeq 106.75,\,\gseq\simeq 3.36$ are the effective relativistic degrees of freedom at the end of reheating and at matter-radiation equality, respectively,
while $\gsp = 2$
denotes its present-day value.
The present-day fractional radiation energy density is $h^2 \Omega_{\rm R}\simeq 4.3\times10^{-5}$. Thus, the resulting GW spectrum encodes both the anisotropic source dynamics during reheating and the modification of tensor-mode propagation due to the non-instantaneous reheating scenario.

\subsection{ Numerical Setup and Diagnostics:}\label{sec:numerical_setup}

To study the dynamics of the radiation and inflaton fields in the reheating phase,
we perform three-dimensional MHD simulations.
%\sout{The governing equations of the system are solved under the open-source {\sc Pencil Code} framework.}
We numerically solve the coupled system of partial differential \Eqq{eq:fluid}
using the \textsc{Pencil Code} \cite{Collaboration2021}
(see \url{https://github.com/pencil-code/pencil-code/}).
The numerical domain is a triply periodic box of comoving size $\bar L^{3}
\equiv (L H_{\rm end})^{3} = (2\pi)^3$ discretised in a uniform mesh.
In this study, we present results from simulations performed with $N^3 = 256^3$ grid points,
as summarised in table ~\ref{tab:sample_table}.

\begin{table}[h!]
\centering
\small
\setlength{\tabcolsep}{3.0pt}      % tighter than default (6pt)
\renewcommand{\arraystretch}{1.15}

\begin{tabular}{ccccccccccc}
\toprule
\toprule
Run & $\omega_\phi$ & $\bar{\Gamma}_\phi$ & $k^n$ &
$\bar{k}_p$ & $\bar{k}_c$ & $ A_{\bar{\rho}_{\rm rad}} $ & $A_{\rm M}$ & $C_s$ & $P_m$ \\
\midrule
R1 & --  & $\infty$   & $k^4$ & 10 & 60 & $10^{-5}$ & $10^{-7}$ & 0.5 & 1 \\
R2 & 0.0 & $10^{-7}$ & $k^4$ & 10 & 60 & $10^{-5}$ & $10^{-7}$ & 0.5 & 1 \\
R3 & 0.0 & $10^{-5}$ & $k^4$ & 10 & 60 & $10^{-5}$ & $10^{-7}$ & 0.5 & 1 \\
%R4 & $256^3$ & --  & $\infty$   & $k^4$ & 10 & 60 & $ 1 $ & $10^{-7}$ & 0.5 & 1 \\
R4 & 0.0 & $10^{-7}$ & $k^4$ & 10 & 60 & $10^{-5}$ & $0$      & 0.5 & 1 \\
R5 & 1./3. & $10^{-5}$ & $k^4$ & 10 & 60 & $10^{-5}$ & $10^{-7}$ & 0.5 & 1 \\
R6 & 1./3. & $10^{-7}$ & $k^4$ & 10 & 60 & $10^{-5}$ & $10^{-7}$ & 0.5 & 1 \\
%R7 & $256^3$ & 0.5 & $10^{-5}$ & $k^4$ & 10 & 60 & $10^{-5}$ & $10^{-7}$ & 0.5 & 1 \\
%R8 & $256^3$ & 0.5 & $10^{-7}$ & $k^4$ & 10 & 60 & $10^{-5}$ & $10^{-7}$ & 0.5 & 1 \\
\bottomrule
\bottomrule
\end{tabular}
\caption{Summary of the run parameters.} %\arp{is 0.3 supposed to be $1/3$?} \purple{yes}}
% \arp{values of $\nu$, $\eta$?}\violet{Dynamic value can not be mentioned in the table} \arp{you should give whichever numerical value you use to describe the viscosity, diffusivity, either constant $\nu$ or coefficient $C_s$ instead, depending on the simulation}.}
\label{tab:sample_table}
\end{table}

%\textcolor{magenta}{DM: Following scaling has to be done: $a\rightarrow a/a_{\rm end}, \rho_{\phi} \rightarrow \rho_{\phi}/\rho_{\rm crt}, \rho_{\rm rad} \rightarrow \rho_{\rm rad}/\rho_{\rm crt}, \eta \rightarrow H_{\rm end} \eta, x \rightarrow H_{\rm end} x, B\rightarrow \sqrt{\rho_{\rm crt}} B, \tilde{\eta}_{\rm M} \rightarrow H_{\rm end} \tilde{\eta}_{\rm M}, \Gamma_\phi \rightarrow \Gamma_{\phi}/H_{\rm end} $}

At the onset of reheating, $\eta=\eta_{\rm end} $,  we initialise the radiation energy density with a log-normal distribution. Therefore, the logarithm of the density will follow a normal distribution, ensuring a positive density value in the model. The initial density profile has a mean value of $1.087 \times 10^{-5}$, RMS value of $1.177 \times 10^{-5}$.%, minimum of $9.988\times 10^{-7}$, and maximum of $ 9.756\times 10^{-5}$ \SD{choose which values you want to quote here}
%\sout{finite fluctuation amplitude determined by the Gaussian random field in logarithmic density}. \sout{Log-normal random distributions are used to model density fields, as they yield a positive density value.} 
This distribution will converge to a normal distribution for small fluctuations~\cite{Wang:2011fj, Kayo:2001gu, Coles:1991if, Xavier:2016elr, Repp:2017znp}. In the linear regime, these perturbations evolve independently and preserve the random phases. Moreover, for reheating scenarios with $\wphi<1/3$, the duration of reheating depends weakly on the initial radiation energy density (see the Appendix for details). 
%\arp{this is not explained in the appendix, is it?}). Therefore, we assume that the initial average radiation energy density is $\langle\trhor (\etae)\rangle = 10^{-5} a^4_{\rm end} \, \rhoc$ with Gaussian-distributed fluctuations \arp{this reference seems very off, why is this here? aren't we referring to initial cosmological perturbations? please address the remaining questions here as they are relevant to understand rest of discussion}
%\arp{clarify if this is mean value or Gaussian fluctuations with zero mean and this rms value}, \arp{on top of mean or zero mean? if on top of mean, then what is rms amplitude?}.

The initial comoving magnetic field spectrum is assumed to follow a broken power law, scaling as $k^n$ on large scales ($k \ll k_p$) and as $k^{-m}$ on small scales ($k \gg k_p$), where $k_p$ denotes the spectral peak. Specifically, we consider
\begin{equation}
E_{\rm M}(k) = 2 \frac{\dd \rho_{\rm M}}{\dd k}
= A_{\rm M}
\frac{(k/k_p)^n \,e^{-(k/k_c)^2}}
{\bigl[1 + (k/k_p)^2\bigr]^{(n+m)/2}},
\end{equation}
where
\begin{equation}
\rho_{\rm M}(\eta) = \frac{1}{2} B_{\rm rms}^2(\eta)
= \frac{1}{2}\int_0^\infty \dd k\,E_{\rm M}(k,\eta)
\end{equation}
denotes the magnetic energy density and $A_{\rm M}$ is the corresponding spectral amplitude. The RMS value of the initial non-helical magnetic field is $1.205\times 10^{-7}$.

In our numerical simulations, we adopt a von K\'arm\'an-type spectrum for the magnetic field, which is commonly motivated by turbulent dynamics, as the initial magnetic spectrum with $n=4$ and $m=5/3$. We stress that this choice is not essential to our analysis and is made primarily as a representative initial condition. Other physically motivated spectral shapes can be considered without altering the general framework of our study. The magnetic spectrum is further suppressed by an exponential cutoff for $k \gtrsim k_c$. The numerical values of the dimensionless characteristic scales, $\bar{k}_p = k_p/\ke$ and $\bar{k}_c = k_c/\ke$, are summarised in Table~\ref{tab:sample_table}, where
$\ke =\mH_{\rm end}$
denotes the comoving wavenumber corresponding to the horizon scale at the end of inflation.
%\color{blue}
The physical value of the horizon wavenumber at the end of inflation depends on the inflationary energy scale and reheating parameters.
It can be expressed as~\cite{ Maiti:2025cbi}
\begin{align}\label{eq:ke}
    \ke^{0} = \left( \frac{43\,\gsre}{11} \right)^{1/3}
    \left( \frac{\pi^2 \gsre}{90} \right)^{\alpha}
    \frac{\HI^{1 - 2\alpha}\, \Tre^{4\alpha - 1} \, T_0}{\Mp^{2\alpha}},
\end{align}
where $\alpha=\frac{1}{3(1 + \wphi)}$, $T_0 \simeq 2.725~\mathrm{K}$ is the present-day CMB temperature, and $\gsre = 106.75$. % \sout{denotes the effective number of relativistic degrees of freedom at the end of reheating.} \arp{new text looks good to me, $\gsre$ has been defined 3 times above already}

%\color{black}
During reheating, the relativistic charged particles produced from the inflaton decay are assumed to be instantaneously thermalised and hence generate a finite electrical conductivity in the plasma.
At temperatures above the electroweak symmetry breaking,
$T>T_{\rm EQ}\simeq100\,\GeV$, the conductivity
of the SM plasma can be approximately expressed as
$\sigma (T) \sim T/(g^2 \ln g^{-1})$, where $g$ is the hypercharge
coupling constant \cite{PhysRevD.56.5254, Arnold:2000dr}.
As an example for the %\sout{highest background} 
reheating
%\arp{I think this should be the relevant as only at later times $\eta_{\rm M}$ will arise}
temperature, i.e.~%\sout{$T_{\rm max}\sim 10^{15}\,\GeV$} 
$\Tre \sim 10^{12}\,\GeV$,
%\sout{implying $\sigma(T_{\rm max})\sim 10^{17}\,\GeV$, and thereforean extremely small}
the fraction of the
magnetic diffusivity to the Hubble time 
is extremely small,
$\eta_{\rm M} H=\sigma^{-1} H
\simeq 2 \times 10^{-8}$ \cite{Durrer:2013pga,RoperPol:2025lgc}.
During the reheating phase, we do not know the exact behaviour of the conductivity, so we assume resistive MHD
with large conductivity, taking the smallest values of $\eta_{\rm M}$ that allow us to perform a numerical simulation for the given discretisation.

In our simulation, we use Smagorinsky subgrid-scale (SGS) modelling to mimic the behaviour of a turbulent plasma with larger Reynolds numbers. The viscosity is proportional to the strength of the rate of strain tensor. This approach allows us to capture the characteristic behaviour expected at high Reynolds numbers without explicitly resolving the physical dissipation scale \cite{Haugen:2004cq, Smagorinsky1963MWRv, Hughes:2001VMS}. For the Smagorinsky SGS model we use $\tilde{\nu} = \nu_s$, where
\begin{align}
    \nu_s = (C_s \Delta)^2 \sqrt{2 S_{ij} S_{ij} }\,.
\end{align}
$S_{ij}$ is the traceless rate-of-strain tensor, $C_s$ is the Smagorinsky constant that corresponds to the average size of the sub-grid-scale eddies, and $\Delta \equiv L /N$ denotes the grid scale of the numerical mesh. We choose $C_s = 0.5$ in our model to suppress any accumulation of kinetic energy density near the Nyquist wavenumber, a phenomenon known as the bottleneck effect in numerical modelling of turbulence.
\section{Results}

\begin{figure*}
    \centering
    \includegraphics[width=0.47\linewidth]{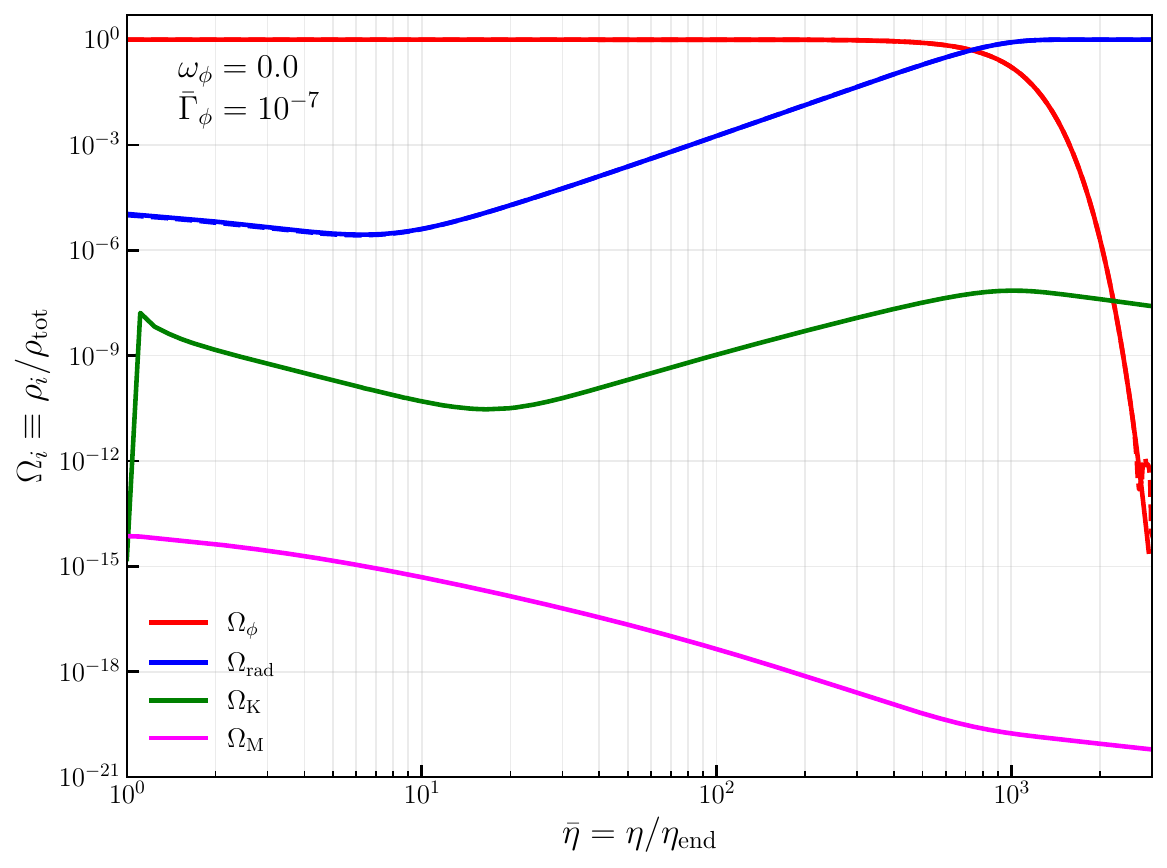}
    \includegraphics[width=0.47\linewidth]{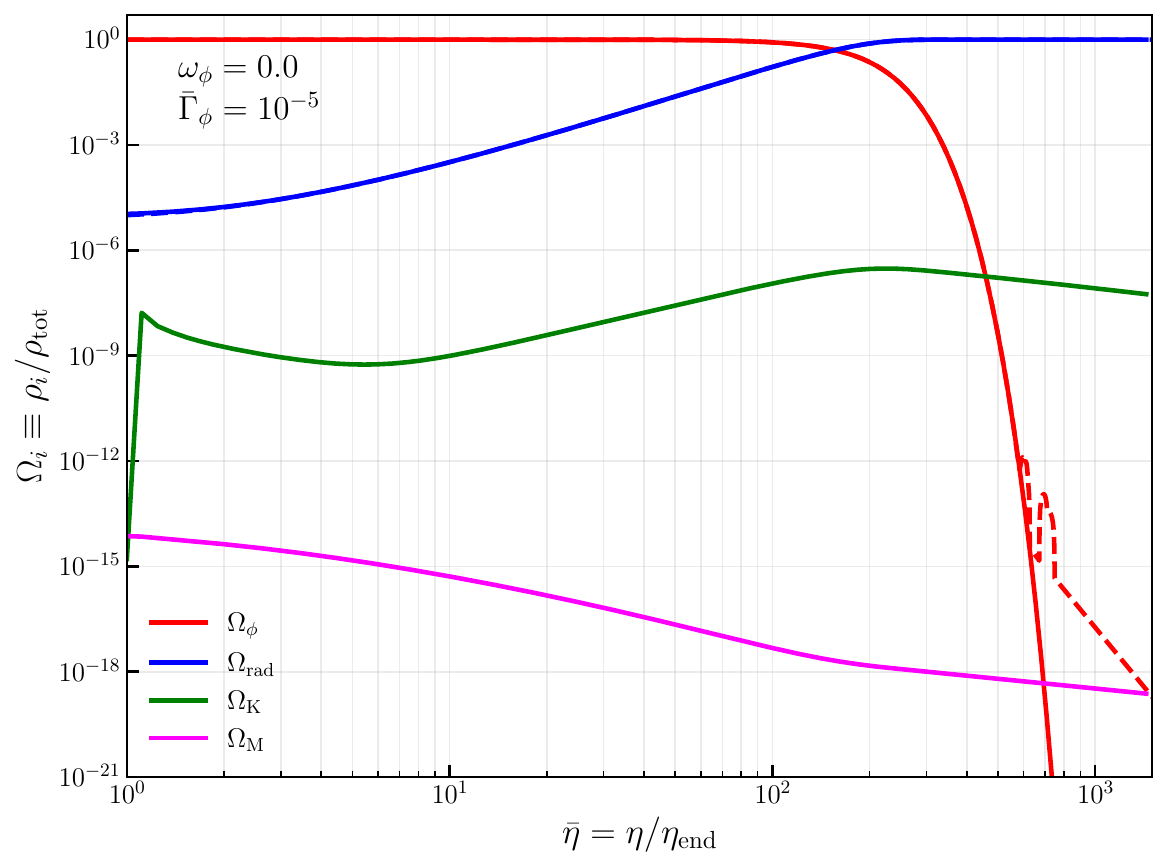}
    \caption{Evolution of the average energy densities of different components normalised by the critical density, i.e.~$\Omega_i=\rho_i/\rhoc$
    for $i = \{$rad, $\phi$, K, M\},
    as a function of conformal time normalised to the initial time
    of reheating,
    i.e. the end of inflation,
    $\bar \eta \equiv \eta/\etae$. The left and right panels
    correspond to the simulations with
    $\bar \Gamma_\phi = 10^{-7}$ and $10^{-5}$, respectively, listed in Table~\ref{tab:sample_table}. }
    \label{fig:energy_density}
\end{figure*}

In this section, we investigate how reheating affects the nonlinear evolution of the primordial plasma. During non-instantaneous reheating, the inflaton gradually decays into radiation, while the resulting plasma develops inhomogeneous velocity and magnetic fields. Figure~\ref{fig:energy_density} shows the evolution of the normalised average energy densities, $\Omega_i (\eta)=\rho_i (\eta)/\rhoc (\eta)$, where the index \enquote{i} denotes the inflaton ($\rho_{\phi}$), radiation ($\rho_{\rm rad}$), kinetic $(\rho_{\rm K} \equiv \tfrac{4}{3} \langle{\rhor \gamma^2 \vur^2\rangle})$, and magnetic $(\rho_{\rm M})$ components. The dashed curves represent the background inflaton and radiation energy densities obtained by solving \Eqq{eq:back_evo} without including inhomogeneities, whereas the solid curves show the corresponding results from the fully nonlinear evolution of the inhomogeneous plasma using the equations~\Eqq{eq:fluid} within the {\sc Pencil Code} framework. The close agreement between the two sets of curves demonstrates that the background evolution in the three-dimensional MHD simulations consistently reproduces the independently calculated background dynamics. The left and right panels correspond to $\bar \Gamma_\phi =10^{-7}$ and $\bar \Gamma_\phi =10^{-5}$, respectively. As expected, a larger inflaton decay rate leads to a shorter reheating period.

At the beginning of the evolution, the inflaton dominates the total energy density, with $\Omega_{\phi}\simeq 1$, while the radiation component is subdominant. During reheating, the inflaton energy density $\Ophi$ approximately follows \Eq{eq:phi_evo}, with its decay becoming significant when the condition $\tGp\simeq {\teta}^{-\frac{3(1+\wphi)}{1+3\wphi}}$ is satisfied. Before radiation domination is established, the inflaton and radiation energy densities follow different scaling laws. For $\eta\ll\ere$, the inflaton energy density evolves as
\begin{align}
\rhop(\eta\ll\ere) \simeq & \tilde \rho_{\phi}^{\rm end} \biggl(\frac{a}{\ae}\biggr)^{-3 (1 + \wphi)}
\nonumber \\ = & \tilde \rho_{\phi}^{\rm end}\, \bar \eta^{-\frac{6(1+\wphi)}{1+3\wphi}},
\end{align}
while the radiation energy density is given by
\begin{align}
\rhor(\eta\ll\ere) =& a^{-4} \trhor \nonumber \\  \simeq & \frac{1 + 3\wphi}
{5 - 3 \wphi} \tGp \tilde \rho_{\phi}^{\rm end} \, \bar \eta^{-\frac{3(1+\wphi)}{1+3\wphi}}.
\end{align}
The radiation energy density therefore decreases more slowly than the inflaton energy density. Consequently, its fractional contribution to the total energy density grows with conformal time as
\begin{equation}
\Ora (\eta \ll \ere) \equiv \frac{\rhor}{\rhoc} \simeq \frac{\rhor}{\rhop} \propto \eta^{\frac{3(1+\wphi)}{1+3\wphi}}.
\end{equation}
For a matter-like reheating phase, this reduces to $\Ora \propto \eta^3$, consistent with the behaviour shown in Fig.~\ref{fig:energy_density}. Reheating ends when the radiation and inflaton energy densities become comparable, at approximately. 
%\purple{Mention if there is any assumption and match with the numerical result}
\begin{align}
\teta\simeq \l(\frac{5-3\wphi}{(1+3\wphi)\tGp}\r)^{\frac{1+3\wphi}{5-3\wphi}}.
\end{align}
Beyond this transition, radiation becomes the dominant component and the normalised radiation energy density approaches $\Ora\simeq 1$.

The nonlinear response of the plasma is reflected in the evolution of the radiation velocity field. We initialise the velocity perturbations as Gaussian random fluctuations with a root-mean-square (rms) amplitude of $u_{\rm rms}\sim10^{-5}$. As shown in Fig.~\ref{fig:energy_density}, the velocity amplitude grows rapidly during the initial stage, reaching $u_{\rm rms}\sim5\times10^{-2}$ within $\eta/\eta_{\rm end}-1\sim10^{-1}$. This early growth is associated with the acceleration of the plasma by pressure gradients. As the radiation component builds up during reheating, its normalised energy density increases from $\sim10^{-5}$ to order unity. The initial rapid growth of the velocity field then subsides, and $u_{\rm rms}$ gradually decreases as the plasma evolves and dissipative effects become important.

The kinetic energy density, $\rho_{\rm K}$, follows a related evolution. Its initial enhancement is associated with the rapid growth of the velocity fluctuations, while its subsequent evolution also reflects the changing radiation density. The normalised kinetic energy density increases from $\sim10^{-15}$ to $\sim10^{-8}$ during reheating. Once radiation domination is established, it gradually decreases, consistent with the decay of the velocity fluctuations and the dissipation of kinetic energy.

The magnetic sector, however, exhibits a markedly different behaviour. The normalised magnetic energy density, $\Omega_{\rm M} \equiv \rho_{\rm M}/\rhoc$, remains nearly constant during the early evolution and decreases at later times, staying strongly subdominant throughout reheating. Thus, despite the substantial enhancement of the inhomogeneous velocity field, the simulations show no corresponding growth of magnetic energy. Within the duration and parameter range explored here, reheating therefore does not produce an efficient transfer of energy into the magnetic sector or lead to an effective dynamo mechanism.

\begin{figure*}
    \centering
    \includegraphics[width=0.48\linewidth]{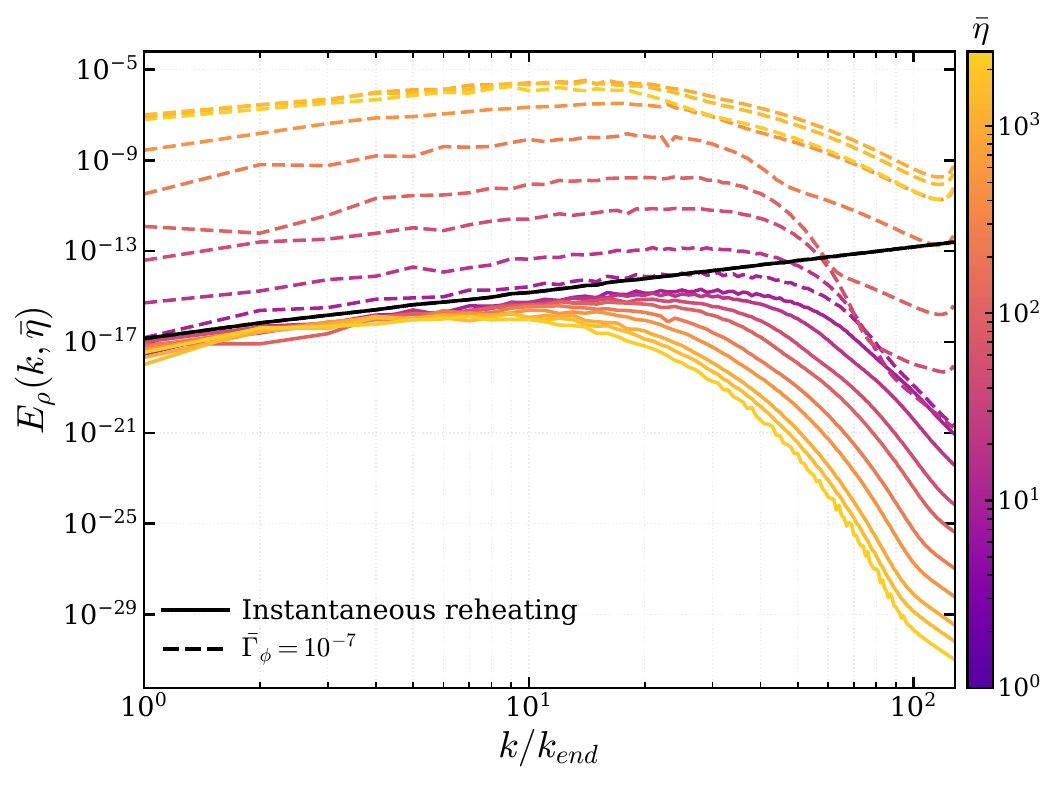}
    \includegraphics[width=0.48\linewidth]{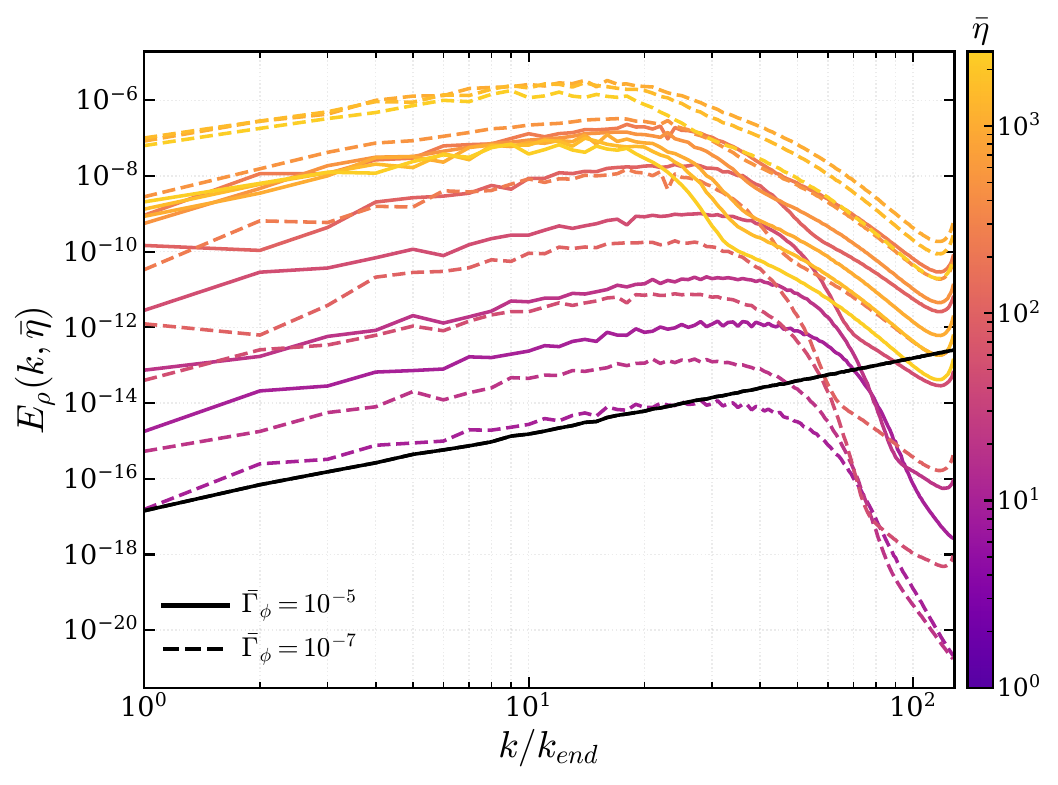}
    \caption{Evolution of the spectrum of the radiation energy density field, $\tilde \rho_{\rm rad}$. In the left panel, we compare the energy spectrum for the scenario of non-instantaneous reheating with $\bar \Gamma_\phi = 10^{-7}$, with the instantaneous reheating scenario. In the right panel, we compare the energy spectrum for two different decay rates, $\bar \Gamma_\phi =10^{-5}$ (solid lines) and $\bar \Gamma_\phi=10^{-7}$ (dashed lines), for $\wphi=0$.}
    \label{fig:density_spectrum}
\end{figure*}

\subsection{Evolution of the radiation density}

In the present homogeneous reheating scenario, we have assumed the inflaton condensate is spatially homogeneous, and the inflaton velocity tends to zero. Since the energy density of the inflaton is spatially homogeneous, the source term in the radiation continuity equation, $\mathcal{S}_r(\eta)$, is also spatially homogeneous and depends only on conformal time. During reheating, the inflaton continuously transfers energy to the radiation plasma. Since the inflaton is homogeneous in our setup, this energy is directly transferred to the homogeneous $k=0$ mode of the radiation density and does not directly source $k>0$ modes. So, the increase in the radiation density increases the power in the $k=0$ mode. Therefore, the mean radiation energy density increases with time, but the interactions with the inflaton do not yield any spatial inhomogeneities. The spatial inhomogeneities of the radiation density arise from the subsequent fluid dynamics.
%, through the advection of the density field and local compression and expansion of the velocity field.
%\SD{Does it sound better, "Therefore the spatial inhomogeneities of the radiation density or variation in the spectrum is result of the advection of the density field and local compression or expansion of the flow field"}

The non-zero $k$ density modes evolve due to the local compression or expansion of the radiation plasma, $\nabla\cdot\vur$, the advection of logarithmic radiation density, $\vur\cdot\nabla\ln\trhor$, Ohmic dissipation, $\tilde\eta_{\rm M}\tJ^2/\trhor$, and work done by the Lorentz force, $\vur\cdot(\tJ\times\tB)/\trhor$, as described in \Eq{eq:3b}.
%The finite-$k$ density modes evolve due to the local compression or expansion of the radiation plasma ($-\frac{4}{3}\nabla\cdot\vur$), advection of of logarithmic radiation density ($-\frac{2}{3}\vur\cdot\nabla\ln\trhor$), Ohmic dissipation ($\tilde\eta_{\rm M}\tJ^2/\trhor$), the Lorentz-force contribution ($-\vur\cdot(\tJ\times\tB)/\trhor$), and the source term $\mathcal{S}_r(\eta)$, as described in \Eq{eq:3b}.
These processes can modify the amplitude of existing density fluctuations and redistribute their power across spatial scales. In particular, regions of compression and expansion can enhance or reduce the local radiation energy density, thereby changing the amplitude of the corresponding finite $k$ modes. The advection term also modifies the spatial variation of the radiation energy density. The electromagnetic contribution to the system is very weak and therefore has a negligible effect on the evolution of the density spectrum.
%For homogeneous reheating scenarios, the ratio of the Lorentz acceleration to the pressure gradient acceleration is of order $10^{-10}$ in our system.
% The electromagnetic contribution is very weak in our simulations:
% at $t=t_{\rm ref}$, the ratio
% \begin{equation}
% \frac{\left|\tilde\eta_{\rm M}\tJ^2-\vur\cdot(\tJ\times\tB)\right|}
% {\left|-\frac{4}{3}\left(\nabla\cdot\vur
% +\frac{1}{2}\vur\cdot\nabla\ln\trhor\right)\right|}
% \sim 10^{-N},
% \end{equation}
%\purple {Numerical estimation, numerical value for reference} 
The spatial inhomogeneities of the radiation density therefore evolve under the combined effects of the advection and compression of the radiation plasma, while the homogeneous radiation background directly increases due to the inflaton source contribution.`

During homogeneous reheating for an equation of state $\wphi=0$, the radiation energy density increases as reheating proceeds, while the overall spectral shape remains approximately unchanged.
After reheating ends, the density spectrum begins to decay. Throughout the evolution, the spectral peak gradually shifts towards smaller wavenumbers. This behaviour is evident in the density power spectrum shown in \Fig{fig:density_spectrum}. For a higher equation of state, the radiation energy density spectrum evolves slightly in a different manner compared to $\wphi=0$. For the homogeneous reheating case with $\wphi=1/3$, the spectrum develops an enhancement at small wave numbers, while the energy density at high $k$ is suppressed (see the Appendix for a detailed discussion).

The high-$k$ modes are strongly damped than the low-$k$ modes due to the diffusion at the grid scale. Hence, density fluctuations on small length scales are dissipated efficiently. The low-$k$ modes are less influenced by the same process and can be enhanced through the non-linear plasma evolution. Therefore, the peak of the density spectrum shifts towards lower wave numbers as the system evolves. Thus, the observed spectral evolution is due to the redistribution and scale-dependent damping of existing density fluctuations and not to a direct injection of energy from the homogeneous inflaton to the finite-$k$ modes.

\subsection{Evolution of the velocity field}
\begin{figure*}
    \centering
    \includegraphics[width=0.48\linewidth]{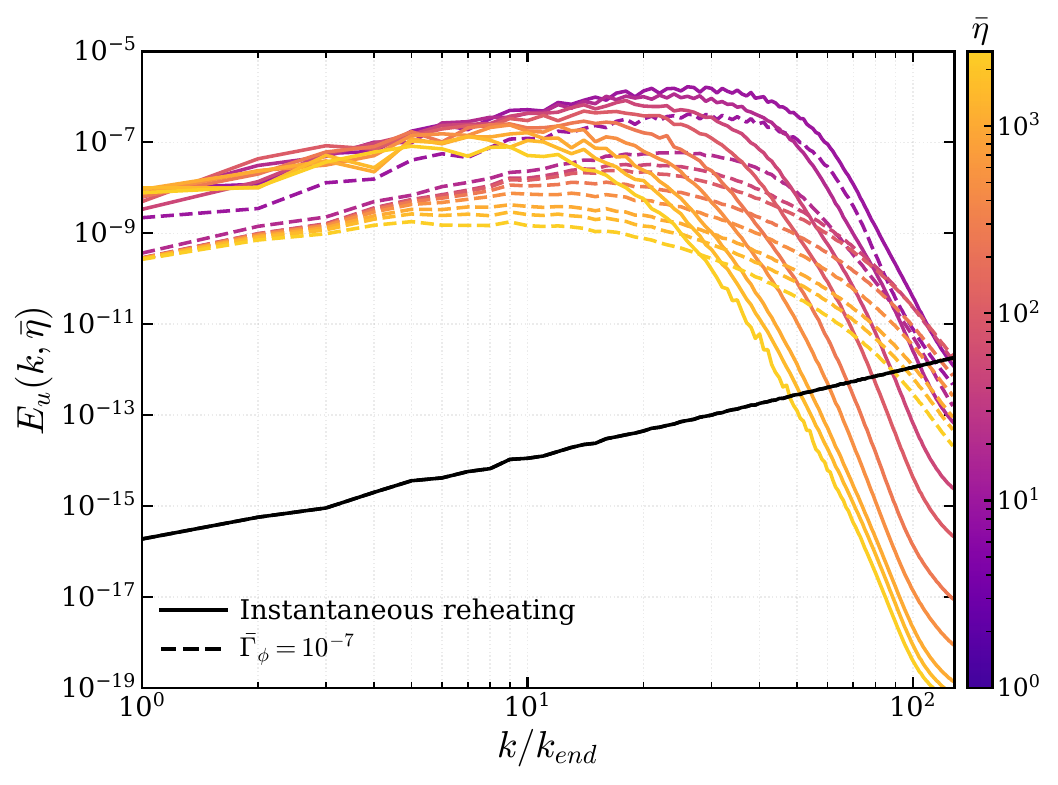}
    \includegraphics[width=0.48\linewidth]{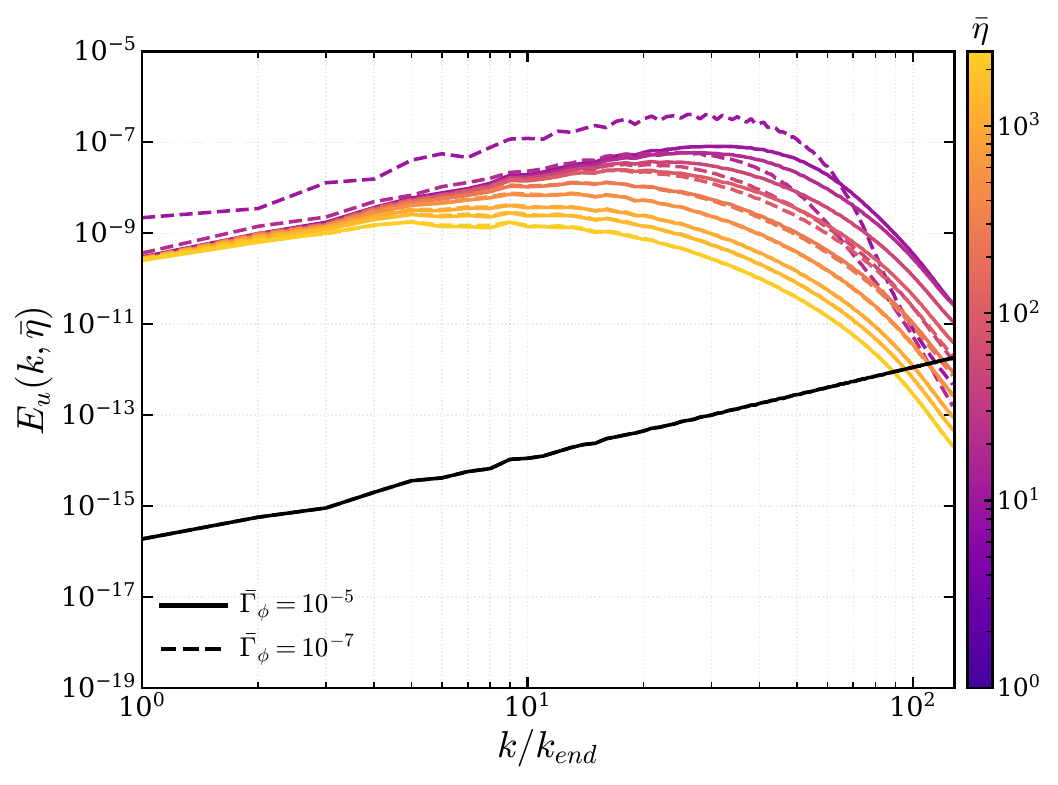}
    \caption{Evolution of the velocity spectrum of the radiation velocity field. In the left panel, we compare the cases with and without reheating. For the reheating case, we take $\tGp=10^{-7}$. In the right panel, we compare the evolution of the velocity spectrum for two different decay rates, $\tGp=10^{-5}$ (solid lines) and $\tGp=10^{-7}$ (dashed lines). In both cases, we consider $\wphi=0$, corresponding to matter-like evolution during reheating.}
    \label{fig:velocity_spectrum}
\end{figure*}

The time evolution of the velocity field shows a sudden increase in the velocity driven by the pressure gradient force, $ \frac{\nabla \tilde{p}}{\tilde{\rho}} $. Following this initial increase, the velocity gradually decreases with time. This behaviour is observed in the velocity power spectrum, shown in \Fig{fig:velocity_spectrum}. In the overall dynamics, the role of the Lorentz force remains negligible due to the weak magnetic field and the associated current density. For homogeneous reheating scenarios, the ratio of the Lorentz acceleration to the pressure gradient acceleration is of order $10^{-10}$ in our system.
%The typical ratio of the Lorentz force and the pressure driven force remains of the order of $10^{-10}$ in homogeneous reheating scenarios.

\begin{figure}
    \centering
    \includegraphics[width=\linewidth]
    %{fig/velocity_nus_vs_etabar.pdf}
    {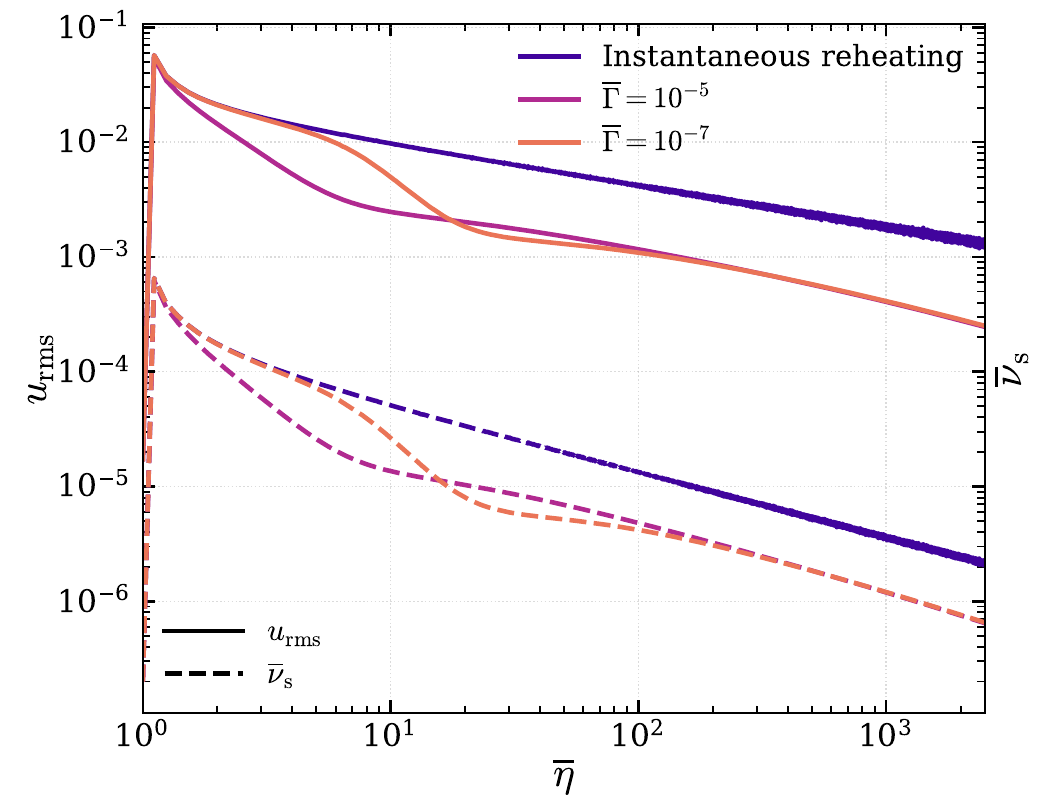}
   \caption{Evolution of the velocity ($u_{rms}$) and SGS viscosity, $\bar{\nu}_{\rm s}$ for instantaneous reheating (solid) and non-instantaneous reheating with $\bar{\Gamma}=10^{-5}$ (dashed) and $\bar{\Gamma}=10^{-7}$ (dotted).}
    \label{fig:viscosity}
\end{figure}

Compared with the instantaneous reheating scenario, the velocity evolution from the non-instantaneous reheating case exhibits a different behaviour due to the sustained turbulence generated in the plasma.  %At the beginning of the evolution, the velocity fluctuation is modified by the pressure gradient.
At the onset of reheating, $\bar{\rho}$ is small, of the order of $10^{-5}$, while $\delta\rho$ is of the order of $10^{-6}$ due to the initial radiation energy distribution. As a result, the relative radiation density perturbation, $ \delta\rho/\bar{\rho} $, is large, leading to a strong pressure gradient force that drives strong flows at the beginning. %That is why we are getting the sudden rise in the velocity at the beginning.
At the early stage of the evolution, the density fluctuation grows by two orders of magnitude, and subsequently, the density fluctuation gradually decreases over time.
%very slowly for instantaneous reheating scenario. 
However, at later stages of non-instantaneous reheating, the mean radiation energy density increases without a significant increase in the inhomogeneities $\delta\rho$.
%Therefore, the relative radiation density perturbation, $\delta\rho/\bar{\rho}$ will decrease over time. So the effect of the pressure gradient on the velocity will also decrease \sout{with time}. 
Therefore, for the homogeneous reheating case, the velocity amplitude exhibits a noticeable decrease at intermediate times. After initially following a behaviour similar to the instantaneous reheating case, $u_{\rm rms}$ drops sharply between $\bar{\eta}\sim5$–$100$ as seen in \Fig{fig:viscosity}. Following this decrease, the velocity continues to decay more gradually.
%This increase in density does not increase the velocity perturbation very much. 
On the other hand, in the instantaneous reheating scenario, the mean radiation density does not increase during the evolution. Consequently, the relative density perturbation, $ \frac{\nabla \tilde{\rho}}{\tilde{\rho}} $, decays gradually with time. As a result, the velocity does not exhibit a pronounced drop but instead decays gradually following the initial jump. Compared to the non-instantaneous scenario, the instantaneous reheating exhibits a larger relative density perturbation. Consequently, the stronger density fluctuations lead to a comparatively larger rms velocity in the instantaneous case.

This behaviour of the velocity field is clearly discernible in the form of power spectrum analysis, as shown in \Fig{fig:velocity_spectrum}. At large scales (small $k$), the velocity in the non-instantaneous reheating scenario is lower than in the instantaneous case. However, at small scales (large $k$), the non-instantaneous scenario retains comparatively more velocity fluctuations. In the instantaneous reheating case, the energy contained at small scales decays much more rapidly than that at large scales.  In contrast, in the non-instantaneous reheating scenario, substantial velocity power at small scales is maintained for a longer period. The difference is evident in the velocity perturbations despite the simulations starting with the same initial conditions and using the same SGS-viscosity model. However, the velocity fluctuation damping at large $k$ depend on the effective SGS viscosity. %SGS viscosity generally damp the velocity spectrum at small scales. 

% The effect of SGS viscosity on velocity evolution  can be written approximately as
% \begin{equation}
% \left(\frac{\partial u_k}{\partial t}\right)_{\rm SGS}
% \sim
% -\nu_{\rm SGS}k^2u_k.
% \end{equation}
% The corresponding damping rate is
% \begin{equation}
% \Gamma_{\rm SGS}\sim\nu_{\rm SGS}k^2.
% \end{equation}

\begin{figure*}
    \centering
    \includegraphics[width=0.32\linewidth]{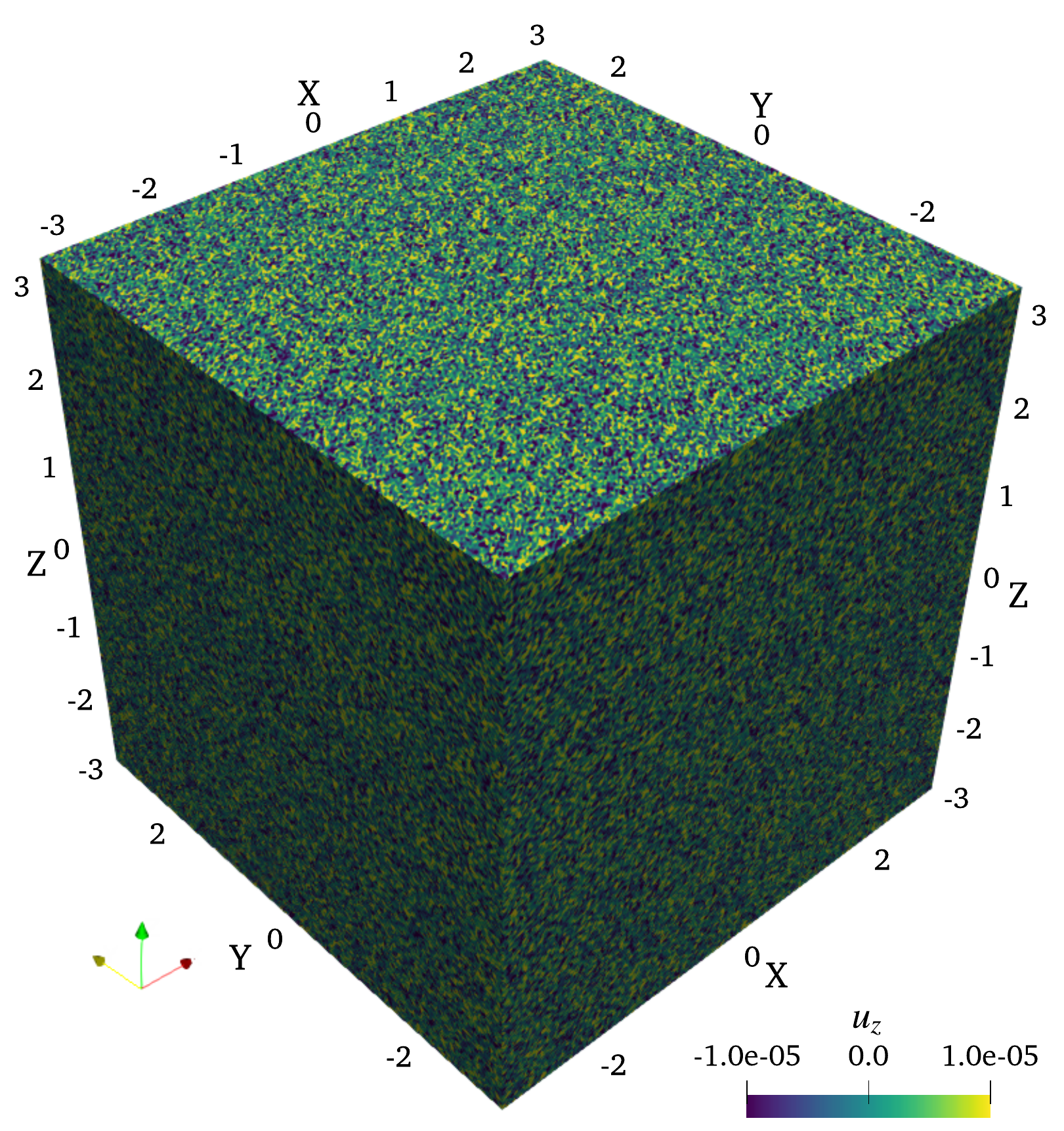}
    \includegraphics[width=0.32\linewidth]{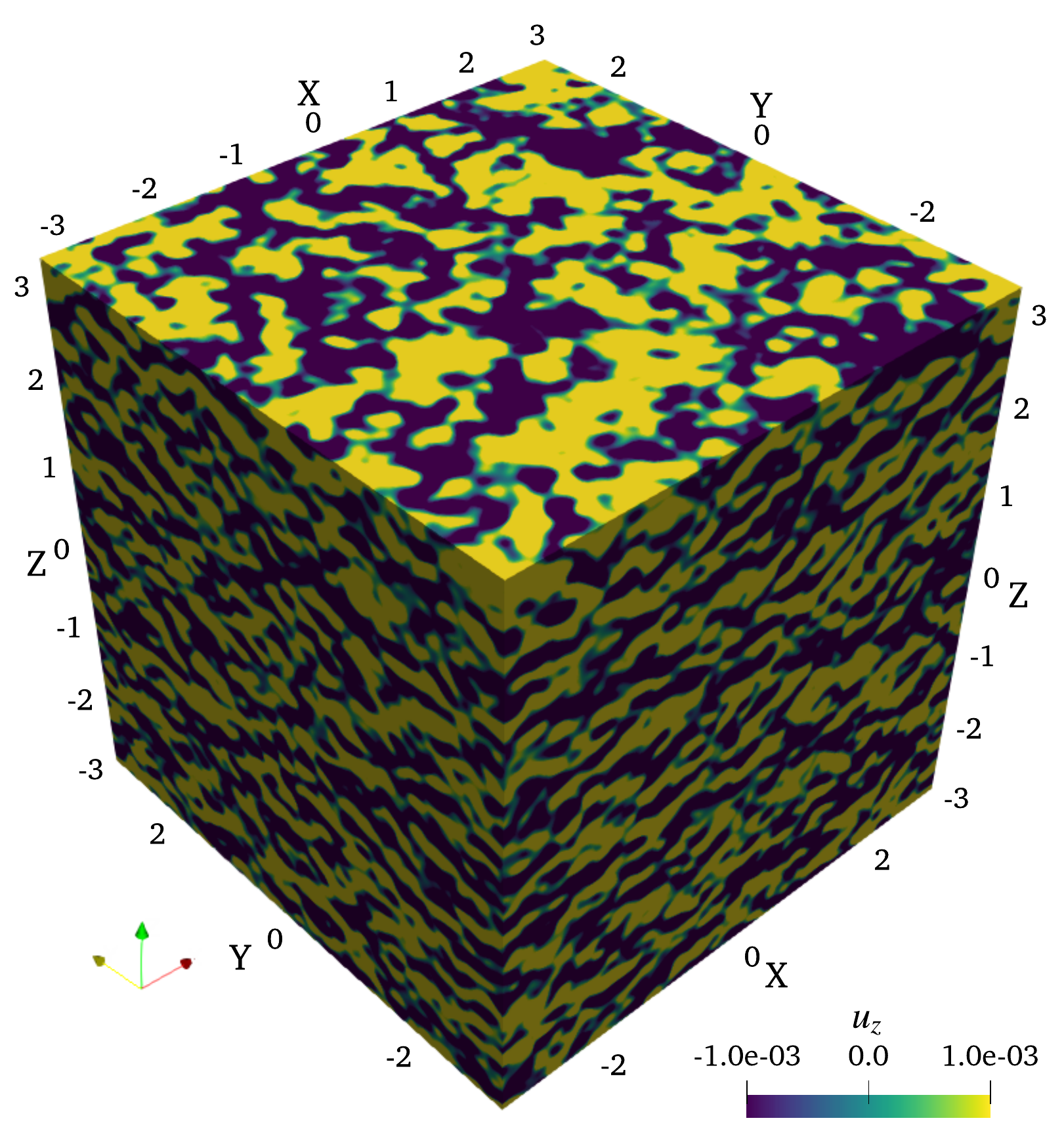}
    \includegraphics[width=0.32\linewidth]{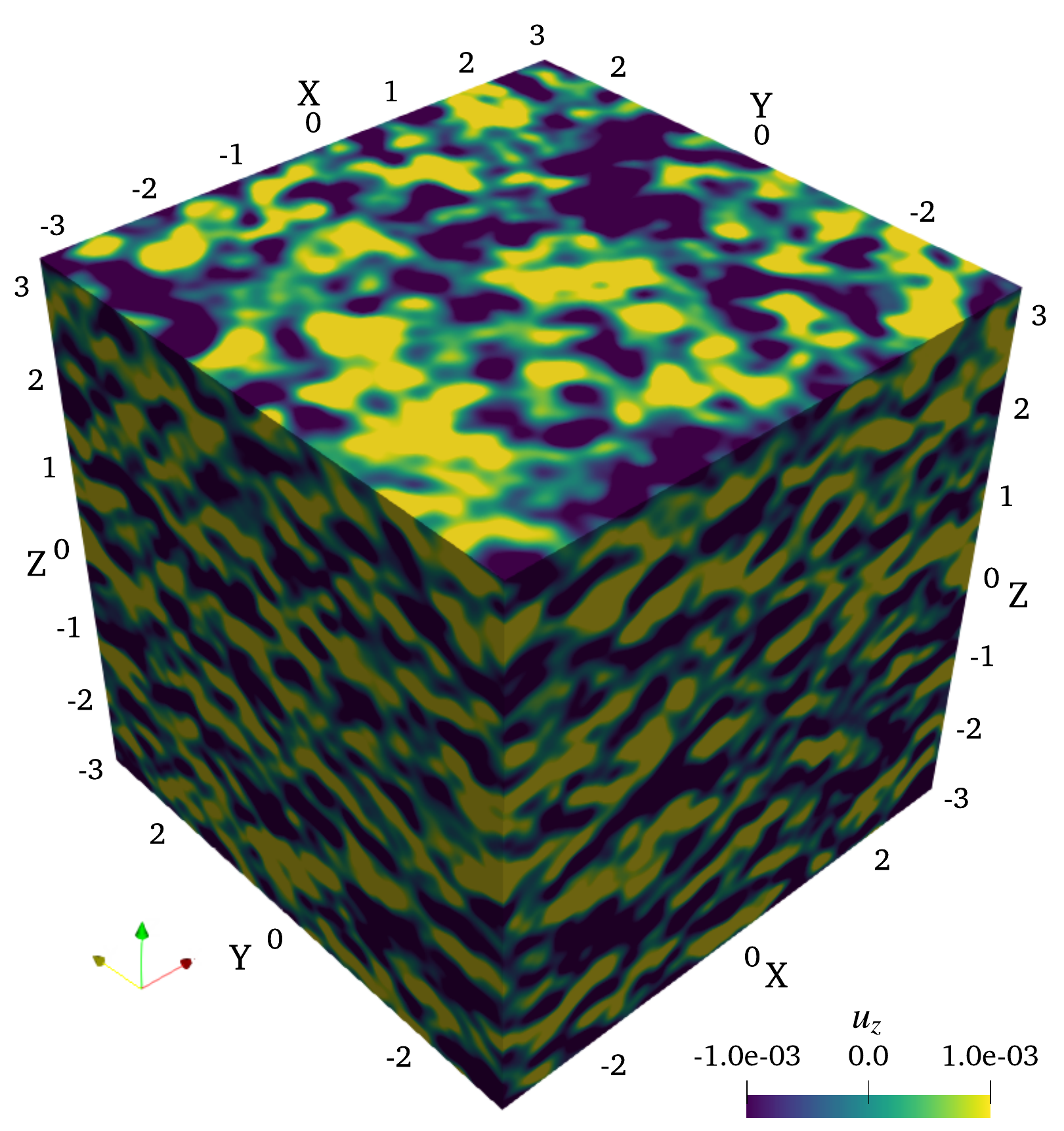} \\
    \includegraphics[width=0.32\linewidth]{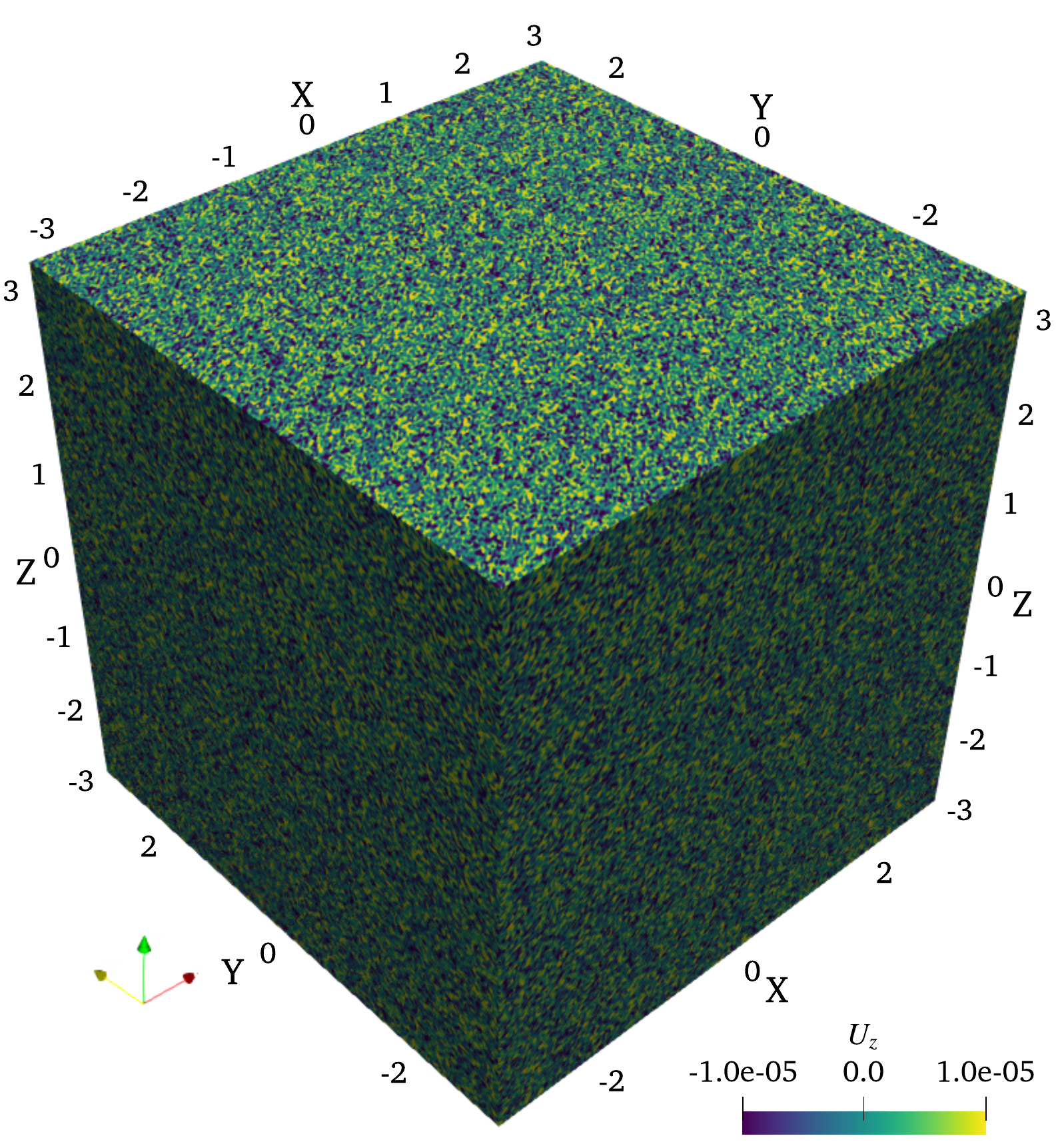}
    \includegraphics[width=0.32\linewidth]{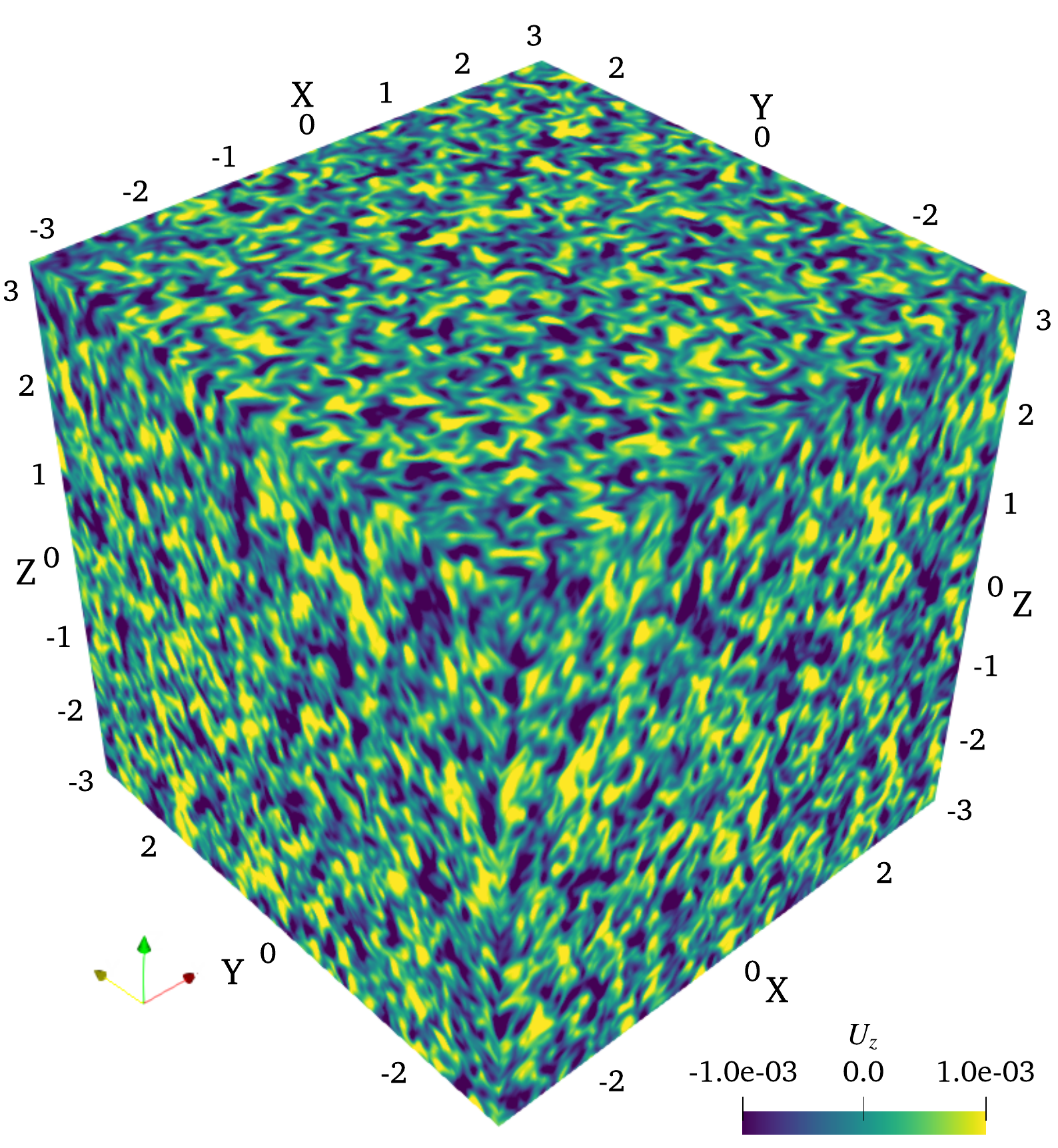}
    \includegraphics[width=0.32\linewidth]{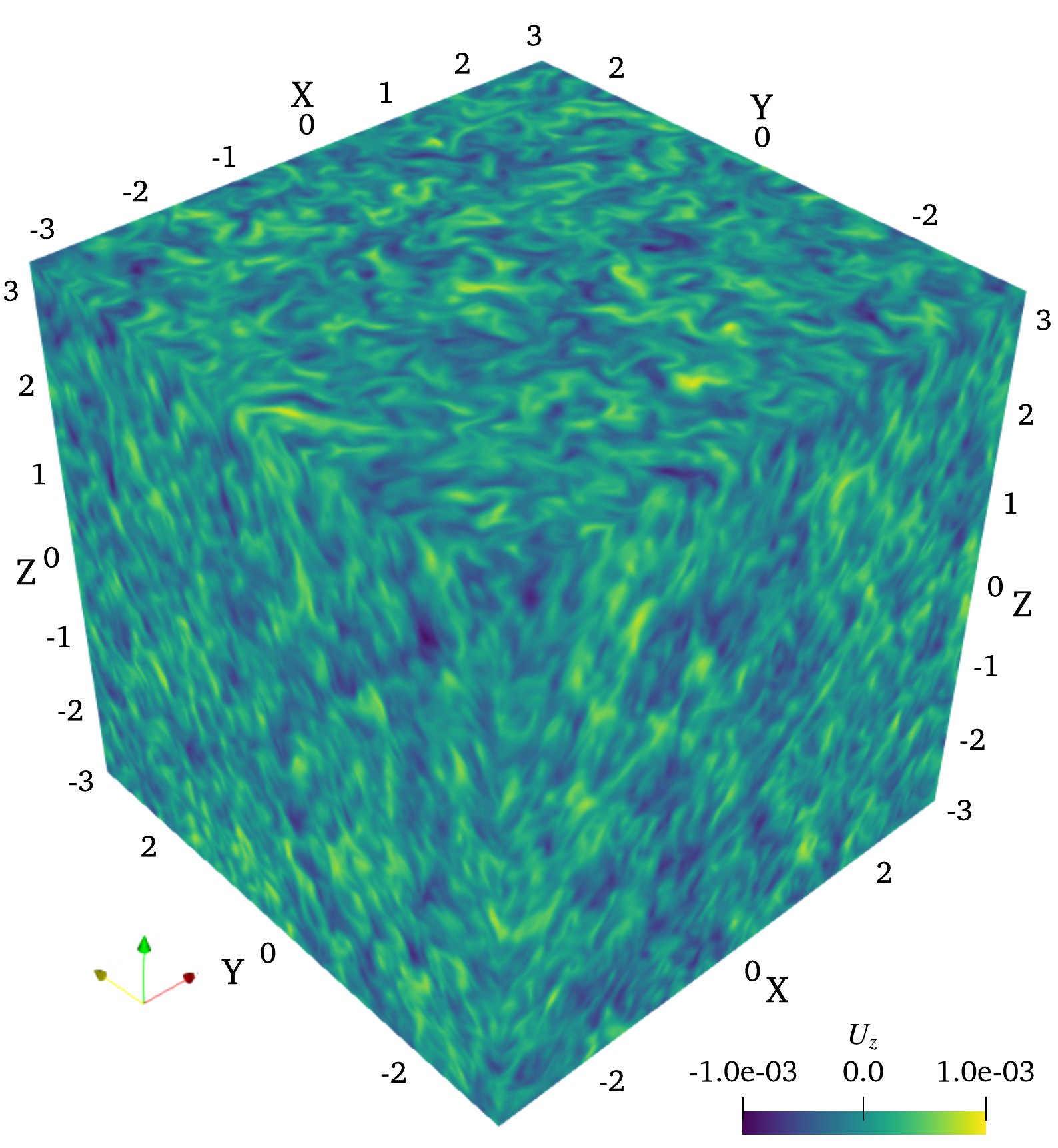}
    \caption{Visualization of $u_z$ at $\bar \eta \equiv \eta/\eta_{\rm end} = 1, 100, 1000$. The colour scale is symmetric about zero. The upper panel shows the velocity evolution for the instantaneous reheating case and  the lower panel shows the evolution of velocity fluctuation for the non-instantaneous reheating ($\tGp=10^{-7}$) case.}
    \label{fig:2D_slices}
\end{figure*}

For homogeneous reheating, the amplitude of the velocity perturbation on large scales is approximately one order of magnitude smaller than in the instantaneous reheating case, which is attributed by the energy injection from the inflaton field. On the contrary, the difference in the velocity amplitude at small scales arises from the variation in the effective viscosity between the two cases. In the instantaneous reheating scenario, viscosity becomes higher self-consistently with a strong rate of strain, leading to enhanced damping in the velocity field (\Fig{fig:viscosity}). 

The evolution of velocity fluctuations is modified in a scale-dependent manner by pressure-gradient and SGS viscosity. In the non-instantaneous reheating case, the continued growth of the mean radiation density alters the pressure-gradient contribution to the velocity field, while the larger SGS viscosity leads to stronger damping of high-wavenumber modes. Consequently, the slower decay of small-scale velocity fluctuations in the non-instantaneous reheating scenario arises from the combined effect of the evolving radiation-density field and SGS dissipation. The different background evolution changes the spectrum of velocity fluctuations that are subject to damping, and the larger SGS viscosity in the instantaneous reheating case further suppresses these fluctuations at small scales. This behaviour is reflected in the velocity-field profiles. In the instantaneous reheating case, we do not observe any turbulent-like flow field, and the characteristic length scale is large (see upper panel of \Fig{fig:2D_slices}). In contrast, distinct filamentary structures dominate with strong small-scale fluctuations in the non-instantaneous reheating case. The z-component velocity field in the 3D simulation box demonstrates intermittent turbulent-like flow structures throughout the evolution (see lower panel of \Fig{fig:2D_slices}).

\subsection{Evolution of vorticity}

\begin{figure}[h]
    \centering
    \includegraphics[width=\linewidth]{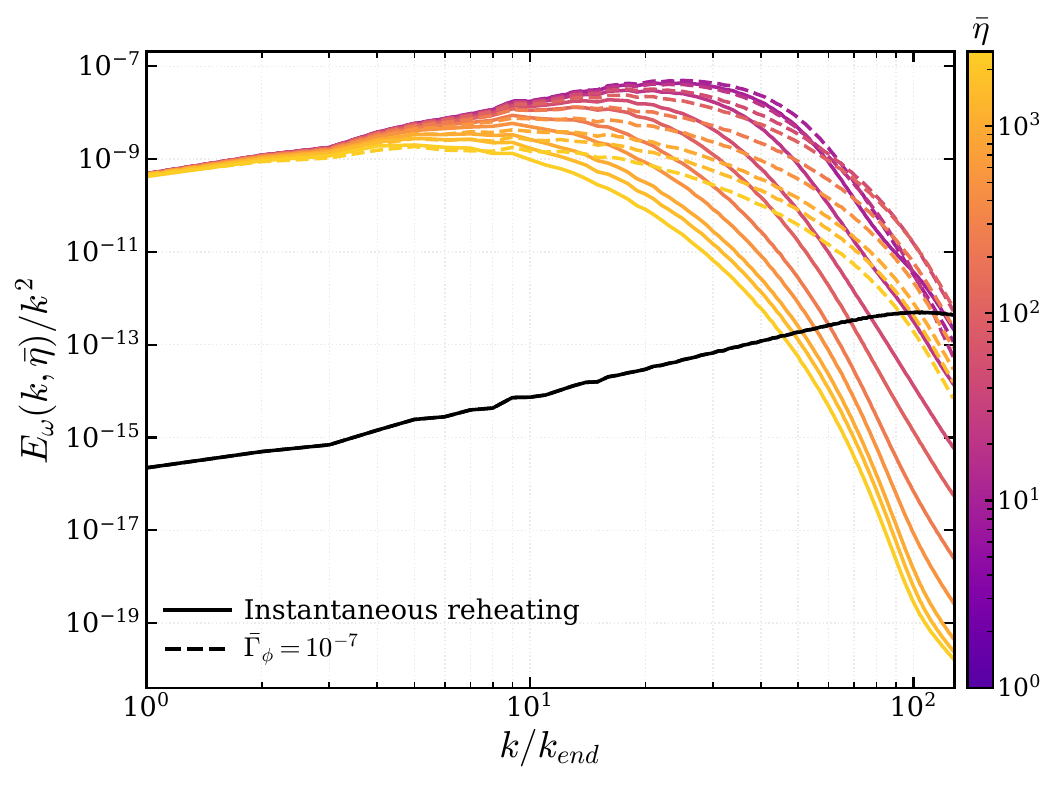}
   \caption{Evolution of the vorticity spectrum with time.
   %\arp{Spectrum of vortical component of velocity, $E_\omega / k^2$ might be more meaningful as one can directly compare it to $E_u$}. 
   The solid lines represent the instantaneous reheating scenario, while the dashed lines correspond to the reheating scenario with a decay rate $\bar \Gamma_\phi = 10^{-7}$. The black solid line denotes the initial vorticity spectrum for both instantaneous and non-instantaneous cases.}
    \label{fig:vorticity_spectrum}
\end{figure}

The vorticity spectrum provides additional evidence for the spatial structure in the velocity field. Vorticity, defined as $\boldsymbol{\omega_r} = \nabla\times\mathbf{u_r}$, characterises the local rotational and shearing motion of the fluid. \Fig{fig:vorticity_spectrum} shows the time evolution of the vorticity spectrum.
%The energy injection from the inflaton during reheating enhances the velocity amplitude at small scales.  
%The velocity field is initialised with spatially variant Gaussian random fluctuations. Since the initial velocity fluctuations have a small but non-zero amplitude, it will produce a non-zero initial seed value for vorticity.
The vorticity evolution equation can be written as \cite{RoperPol:2025lgc},
\begin{align}
\frac{\partial\boldsymbol{\omega_r}}{\partial \eta}
={}&
-\underbrace{(\mathbf{u_r}\cdot\nabla)\boldsymbol{\omega_r}}
_{\substack{\text{Advection}\\\text{term}}}
-\underbrace{\boldsymbol{\omega_r}(\nabla\cdot\mathbf{u_r})}
_{\substack{\text{Compression/}\\\text{expansion}}}
+\underbrace{(\boldsymbol{\omega_r}\cdot\nabla)\mathbf{u_r}}
_{\substack{\text{Vortex}\\\text{stretching}}}
\nonumber\\
&+
\underbrace{
\frac{\boldsymbol{\omega_r}\,\Psi}
{(1- c_s^2 u_r^2)\gamma^2}
}_{\substack{\text{Vorticity}\\\text{correction}}}
+
\underbrace{
\mathbf{u_r}\times\nabla
\left[
\frac{\Psi}
{(1- c_s^2 u_r^2)\gamma^2}
\right]
}_{\substack{\text{Vorticity}\\\text{generation}}}
\nonumber\\
&+
\underbrace{
\frac{
\left(\nabla\tilde{p}_{rad} -\tilde{\mathbf{f}}_{\rm tot}\right)
\times
\nabla\left(\trhor\gamma^2\right)
}
{(1+ c_s^2)\trhor^2\gamma^4}
}_{\substack{\text{Baroclinic}\\\text{and force contribution}}}
+
\underbrace{
\frac{
\nabla\times\tilde{\mathbf{f}}_{\rm tot}
}
{(1+ c_s^2)\trhor\gamma^2}
}_{\substack{\text{External}\\\text{force}}}.
\label{eq:relativistic_vorticity_evolution}
\end{align}

where the scalar quantity $\Psi$ is given by
\begin{align}
\Psi={}&
c_s^2\,\nabla\cdot\mathbf{u_r} + c_s^2\frac{1-c_s^2}{1+c_s^2}(\mathbf{u_r} \cdot\nabla) \ln\trhor -\frac{\tilde{f}^{0}_{\rm tot}}{\trhor}
\nonumber\\
& + \frac{2 c_s^2}{1+ c_s^2} \frac{\mathbf{u}\cdot\tilde{\mathbf{f}}_{\rm tot}}
{\trhor} + (3 c_s^2-1)\mathcal{H}
\end{align}

and $\tilde{f}^{\mu}_{\rm tot}$ denotes any external force, e.g., the Lorentz force and viscous force,
\begin{equation}
\tilde{\mathbf{f}}_{\rm tot} = \tilde{\mathbf{f}}_{\rm Lor} + \tilde{\mathbf{f}}_{\rm visc}
\end{equation}

For a radiation plasma, $c_s^2 = 1/3 $ and $\tilde{p} = \tilde{\rho}/3$. Therefore, the baroclinic torque becomes zero. 
%Besides, for an inflaton equation of state $\omega_\phi = 0$, the vorticity generation term associated with $\omega_\phi$ is also eliminated, and 
So, the production of vorticity is not directly modified by the radiation density. The velocity field is initialised with spatially varying Gaussian random fluctuations. Since the initial velocity fluctuations have a small but non-zero amplitude, they produce a non-zero initial seed value for vorticity. This initial vortical fluctuation in the radiation plasma will be modified by advection and compression, vortex stretching and tilting, as well as by the Lorentz force and dissipative terms present in the governing equation. In our system, the magnetic field and the current density remain weak. Therefore, in the absence of a strong Lorentz force, vorticity evolution is dominated by advection and compression, including the vortex stretching and tilting sources.
%However, for an inflaton equation of state $\omega_\phi = 1/3$, the vorticity generation term will survive. Therefore, for $\omega_\phi = 1/3$, the vorticity generation term will also contribute to the generation of vorticity. Thus, even if the initial vorticity in the system is zero, the system can generate vorticity.

At the onset of reheating, we find a sudden increase in the vorticity spectra at large scales by almost six orders of magnitude (\Fig{fig:vorticity_spectrum}). This enhancement is a direct consequence of the increase in the velocity field associated with compression, vortex-stretching, and tilting terms.
Following this initial enhancement, the two reheating scenarios exhibit different behaviour at large wavenumbers.  For instantaneous reheating, the amplitude of $E_\omega(k,\bar{\eta})/k^2$ decays with time for large $k$, and the spectral peak shifts towards smaller $k$ gradually. In contrast, the non-instantaneous reheating scenario retains substantially more power at large wavenumbers.  The difference between the two scenarios becomes particularly pronounced at late times. This behaviour is consistent with the continuous transfer of energy from the inflaton to the radiation fluid during reheating that delays the decay of the associated vortical motions. However, the small-scale evolution is also affected by the SGS viscosity used in our simulations, which adds more dissipation to velocity gradients and suppresses the high-k modes. We thus find that the slower decay of vorticity at small scales in the homogeneous reheating scenario is due to the combined effect of 
the energy injection from the inflaton field and the SGS dissipation.

% Small scale nature why vorticity is decaying more

%It is clearly visible in the vorticity spectrum. 
%The amplitude at large scales grows by about six orders of magnitude.

%we can seee enhancenment in the amplitude of the vorticity we can   see the \Fig{fig:vorticity_spectrum}

%also see that the vorticity is enhanced at small scales for the non-instantaneous reheating scenario.  This behaviour is consistent with the observed small-scale structure in the velocity field %and the kinetic energy spectrum.  The persistent rotation in the velocity field indicates that the energy injection during reheating continuously drives the turbulence. So these small-scale vortical modes will not decay throughout the reheating period. 
%As a result, the radiation plasma will be in a nonlinear state during the reheating period and helps to sustain the turbulence in the plasma.

% \paragraph{Evolution of the magnetic field:}

\subsection{Evolution of the magnetic field}
%\color{blue}
In the absence of a direct coupling between the inflaton and gauge sectors, the magnetic field evolves solely through the plasma dynamics. In the ideal-MHD limit, %appropriate for the large magnetic Reynolds numbers considered here, 
the induction equation is
\begin{equation}
\frac{\partial \mathbf{\tB}}{\partial \eta}
=
\nabla\times(\mathbf{u_r}\times\mathbf{\tB}),
\end{equation}
or, equivalently,
\begin{equation}
\frac{\partial\mathbf{\tB}}{\partial\eta}
=
(\mathbf{\tB}\cdot\nabla)\mathbf{u_r}
-
(\mathbf{u_r}\cdot\nabla)\mathbf{\tB}
-
\mathbf{\tB}(\nabla\cdot\mathbf{u_r})
\end{equation}
The first term on the right-hand side describes the stretching of magnetic-field lines by velocity gradients and is responsible for dynamo amplification. Efficient amplification therefore requires a sufficiently strong and sustained velocity field with nontrivial spatial structure.

In our reheating scenario, the inflaton decay rate is spatially homogeneous and constant over time. The resulting energy transfer to the radiation component is therefore homogeneous and does not directly generate additional spatial inhomogeneities in the radiation fluid. Although the radiation energy density increases continuously during reheating, this homogeneous energy injection does not efficiently source velocity perturbations \Fig{fig:velocity_spectrum}. Instead, the evolution of the velocity field is governed by the radiation pressure-gradient and nonlinear advection terms in the momentum equation [see \Eq{eq:3d}], which depend on spatial gradients of the radiation density. Since these gradients remain small in the homogeneous reheating setup, the velocity perturbations exhibit no significant growth.

This behaviour is also reflected in the kinetic energy density, $\rho_{\rm K}
=
\frac{4}{3}
\left\langle
\rho_{\rm rad}\gamma^2\vur^2
\right\rangle$.
While $\rho_{\rm K}$ increases during reheating, its evolution primarily tracks the growth of the radiation energy density rather than an increase in the velocity amplitude~\Fig{fig:energy_density}. The velocity perturbations therefore remain approximately unchanged, and the subsequent decay of the velocity field after reheating leads to a corresponding decrease in $\rho_{\rm K}$.

We further characterize the development of small-scale velocity structure through the vorticity and its power spectrum. As shown in \Fig{fig:vorticity_spectrum}, the homogeneous- and instantaneous-reheating scenarios exhibit comparable vorticity amplitudes at large scales. The homogeneous-reheating case shows somewhat more energy densities at small scales, but this does not exhibit sustained growth with time. Hence, the additional small-scale vorticity does not signal the development of a progressively stronger turbulent velocity field, which is consistent with the nearly unchanged velocity-spectrum amplitude.

%SD: add Lorentz force

\begin{figure}
    \centering
    \includegraphics[width=\linewidth]{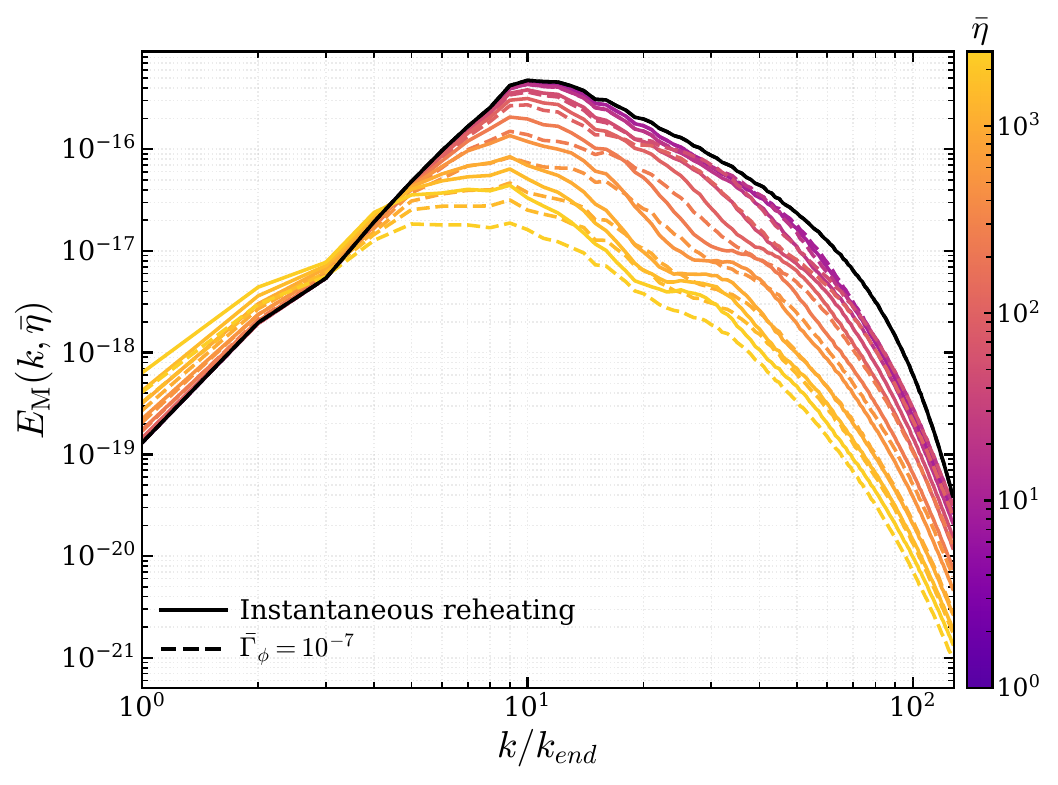}
   \caption{Time evolution of the magnetic spectrum for two scenarios. The solid lines represent the instantaneous reheating case, while the dashed lines correspond to the reheating scenario with decay rate $\bar \Gamma_\phi = 10^{-7}$. The black solid line denotes the initial magnetic spectrum.}
    \label{fig:magnetic_spectrum}
\end{figure}

The absence of sustained growth in the velocity field and its spatial gradients prevents efficient stretching, twisting, and folding of magnetic field lines.
As a result, magnetic fields do not show any signature of the dynamo amplification. In \Fig{fig:magnetic_spectrum}, we present the evolution of the magnetic power spectrum and compare it with the instantaneous-reheating scenario for $\bar{\Gamma}_\phi=10^{-7}$. The magnetic spectrum remains nearly unchanged throughout reheating and closely follows the instantaneous-reheating result, with no modification of its spectral shape. Thus, homogeneous inflaton decay and the associated growth of the radiation background are insufficient, by themselves, to generate additional magnetic field amplification.

This conclusion relies on the homogeneous reheating assumption adopted here. Inhomogeneous reheating can, in principle, provide additional sources for velocity perturbations. For example, a spatially varying inflaton decay rate, $\Gp=\Gp(\vx,\eta)$, would induce inhomogeneous energy transfer to the radiation component and could generate additional density and velocity perturbations. Similarly, coupling the inhomogeneous inflaton velocity field directly to the radiation velocity could provide an additional source of velocity gradients. Such mechanisms may produce a more complex velocity field and potentially enhance magnetic-field amplification through dynamo-like processes~\cite{Calzetta:1997ku, Calzetta:2010zu, Calzetta:2015hda, Iarygina:2026bom}. A detailed study of these inhomogeneous reheating scenarios and their impact on magnetic-field evolution will be studied in our future work.
% \begin{figure*}
%      \includegraphics[width=0.47\linewidth]{fig/comparison_GWs_spectrum.pdf}
%     \includegraphics[width=0.47\linewidth]{fig/comparison_GWs_spectrum_reheating.pdf}
%    \caption{The GW spectra as a function of the normalized wave number for different reheating scenarios. Left panel: comparison between instantaneous reheating (solid lines) and non-standard reheating (dashed lines). Right panel: GW spectra for two decay rates, $\tGp=10^{-5}$ (solid lines) and $\tGp=10^{-7}$ (dashed lines). The effective equation of state is fixed to $\wphi=0.0$ for all reheating scenarios.}
%     \label{fig:GWs_spec}
% \end{figure*}
\begin{figure*}
     \includegraphics[width=0.47\linewidth]{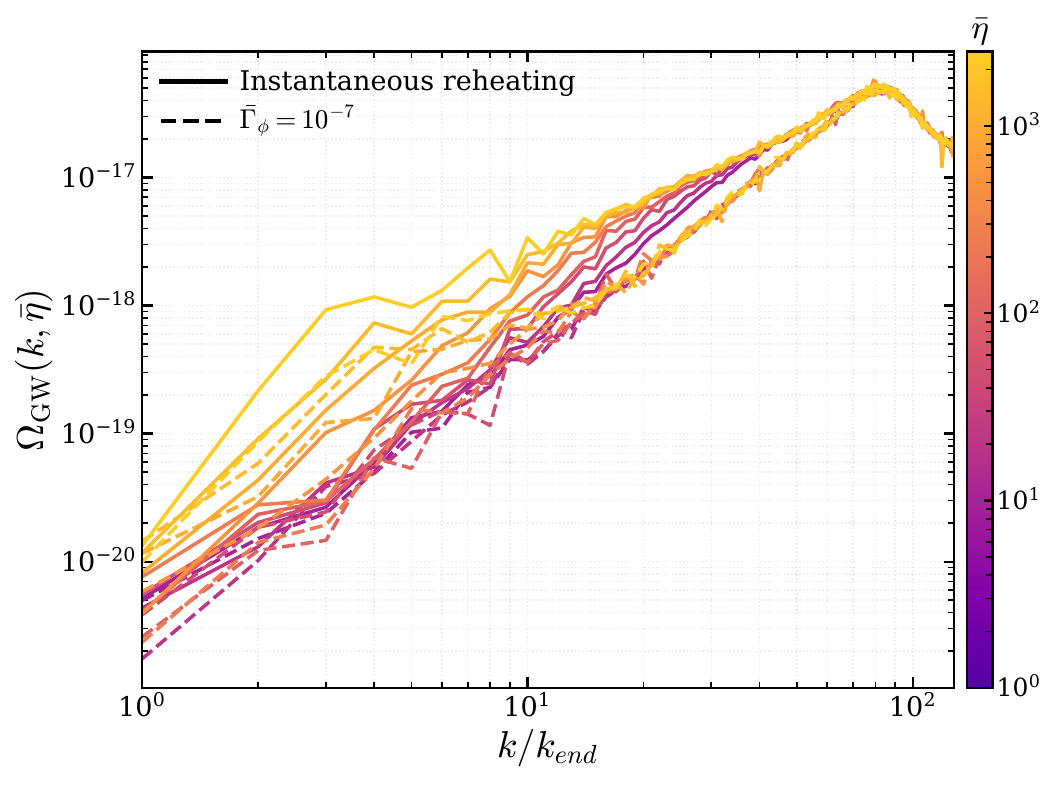}
    \includegraphics[width=0.47\linewidth]{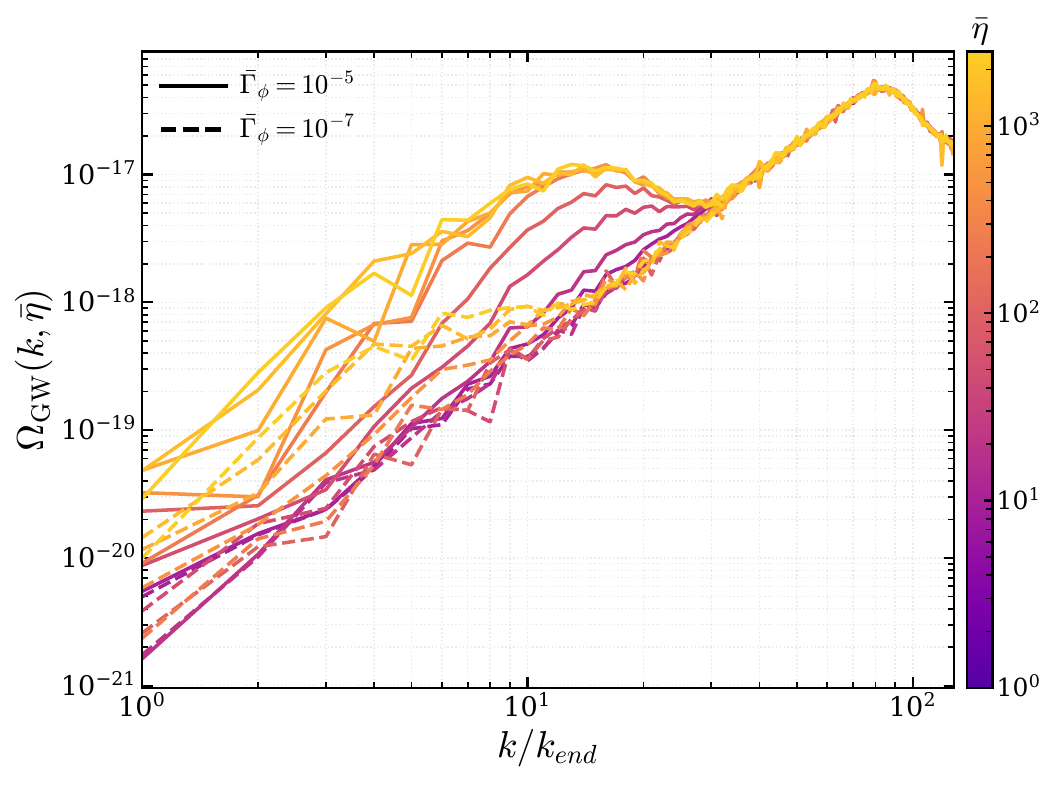}
   \caption{Spectrum of GWs for different reheating scenarios. Left panel: comparison between instantaneous reheating (solid lines) and non-instantaneous reheating (dashed lines). Right panel: GW spectrum for two decay rates, $\bar \Gamma_\phi =10^{-5}$ (solid lines) and $\bar \Gamma_\phi =10^{-7}$ (dashed lines). The effective equation of state is fixed to $\wphi=0$ for all reheating scenarios.}
    \label{fig:GWs_spec}
\end{figure*}

% \paragraph{Gravitational wave production:}

%\color{black}
\subsection{Gravitational wave production}
\begin{comment}

\begin{figure}
    \centering
    \includegraphics[width=\linewidth]{fig/comparison_omega_GWs_spectrum_present_today.pdf}
   \caption{ Spectra of $\Omega_{GW}(k,\eta_0)h^2$ evaluated at at the present day for two decay rates, $\bar \Gamma_\phi =10^{-5}$ (solid lines) and $\bar \Gamma_\phi =10^{-7}$ (dashed lines).}
    \label{fig:presentday_GWs_spec}
\end{figure}
\end{comment}
In this section, we discuss the production and evolution of the GW spectrum and its possible observational signatures in the non-instantaneous reheating scenario. In our framework, GWs can be sourced through two channels: velocity perturbations and the magnetic field. As discussed in the previous section, the magnetic field does not undergo significant amplification in the homogeneous reheating scenario considered here, owing to the absence of sufficiently strong velocity anisotropies and sustained dynamo activity. Consequently, the magnetic contribution to GW production is negligible. The velocity perturbations, however, are affected by the reheating dynamics through their nonlinear evolution and redistribution of energy across different length scales. Since velocity perturbations provide a source for GWs, this scale-dependent evolution can leave a characteristic imprint on the GW spectrum.

% \begin{figure}
%     \centering
%     \includegraphics[width=\linewidth]{fig/comparison_omega_GWs_spectrum_present_day_frequency.pdf}
%    \caption{ Spectra of $\Omega_{GW}(f,\eta_0)h^2$ evaluated at the present day for two decay rates, $\bar \Gamma_\phi =10^{-5}$ (solid lines) and $\bar \Gamma_\phi =10^{-7}$ (dashed lines).}
%     \label{fig:presentday_GWs_spec_freq}
% \end{figure}

\begin{figure}
    \centering
    \includegraphics[width=\linewidth]{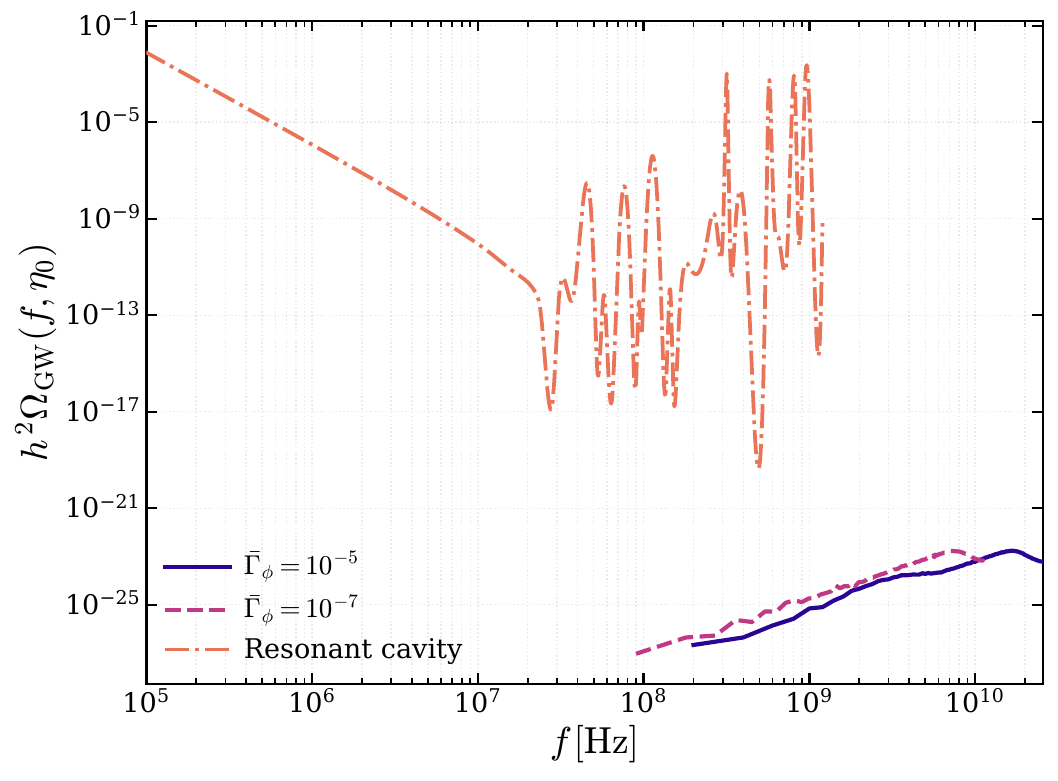}
   \caption{ Spectra of $\Omega_{GW}(f,\eta_0)h^2$ evaluated at the present day for two decay rates, $\bar \Gamma_\phi =10^{-5}$ (solid lines) and $\bar \Gamma_\phi =10^{-7}$ (dashed lines). The reference spectrum for the resonant cavity is shown by the dot-dashed curve.}
    \label{fig:presentday_GWs_spec_freq}
\end{figure}

In Fig.\ref{fig:GWs_spec}, we show the dimensionless GW energy density spectrum, $\Omega_{\rm GW}(k,\eta)$, as a function of the normalised wave number $k/k_{\rm e}$. Using Eq.\eqref{eq:ke}, the initial comoving horizon scale, and hence the characteristic wave number $k_{\rm e}$, can be determined in terms of the inflationary energy scale and the parameters governing reheating. For our benchmark choice, $\wphi=0$ and $\bar{\Gamma}_\phi=10^{-5}$, corresponding to an effective reheating temperature of $T_{\rm re}\sim 10^{12},\GeV$, we obtain $k_{\rm e}\simeq 10^{23},\Mpc^{-1}$ for $H_{\rm I}\simeq 10^{13},\GeV$. This corresponds to a present-day frequency of approximately $f_{\rm e}\simeq 2\times10^{8},\Hz$. Keeping $\wphi=0$ but decreasing the decay rate to $\bar{\Gamma}_\phi=10^{-7}$ shifts the characteristic frequency to $f_{\rm e}\simeq 9\times10^{7},\Hz$, demonstrating that the characteristic scale depends on both the inflationary energy scale and the reheating history.

Fig. \ref{fig:GWs_spec} shows the evolution of the GW spectral energy density during reheating. In the left panel, we compare the instantaneous-reheating case with a prolonged reheating scenario characterised by $\bar{\Gamma}_\phi=10^{-7}$. In the right panel, we compare $\bar{\Gamma}_\phi=10^{-5}$ (solid curves) and $\bar{\Gamma}_\phi=10^{-7}$ (dashed curves), keeping $\wphi=0$ fixed. The resulting spectra exhibit a clear dependence on the reheating history, despite identical underlying inflationary parameters.

This dependence originates from the nonlinear evolution of the velocity field. Initially, the inflaton behaves as a coherent, approximately homogeneous condensate, with most of its energy concentrated in the zero-momentum mode. As the condensate decays into radiation and nonlinear interactions develop, energy is transferred from the infrared sector towards finite wave numbers, generating inhomogeneous field and velocity perturbations over a broad range of scales. The efficiency and timescale of this redistribution depend on the duration of reheating. A smaller decay rate corresponds to a more prolonged reheating epoch, allowing the system to undergo nonlinear evolution and energy redistribution for a longer period before reaching the radiation-dominated regime.

The resulting GW spectrum therefore depends on the evolution of the velocity perturbations during reheating, with the differences between the scenarios being particularly pronounced in the infrared region. The shorter-reheating case exhibits a relatively stronger transfer of power towards finite wave numbers at the times relevant for GW production, leading to an enhanced velocity perturbation spectrum on these scales and, consequently, a larger GW amplitude. For prolonged reheating, the velocity spectrum undergoes further nonlinear evolution and redistribution across scales, resulting in a modified spectral shape and amplitude. Thus, the dependence of the GW signal on $\Gamma_\phi$ arises from its control over the duration of the nonlinear evolution rather than from the decay rate acting as a direct source of GWs. 

The present-day GW spectrum provides a direct illustration of this dependence. In Fig.~\ref{fig:presentday_GWs_spec_freq}, we show $\Omega_{\rm GW,0}(k)$ for $\tGp=10^{-5}$ and $\tGp=10^{-7}$ with $\wphi=0$. Although the overall spectral behaviours are similar, their characteristic peak frequencies are different. For $\bar{\Gamma}_{\phi}=10^{-7}$, the spectrum reaches its maximum at approximately $f_{\rm peak}\sim 7.5\times10^9\,\mathrm{Hz}$, whereas for $\bar{\Gamma}_{\phi}=10^{-5}$, the spectrum reaches its maximum near $f_{\rm peak}\sim 2\times10^{10}\,\mathrm{Hz}$. Thus, increasing the inflaton decay rate shifts the characteristic GW frequency towards higher values. The two spectra exhibit a clear difference in their infrared tails, indicating that the large-scale GW spectrum retains information about the reheating history. In particular, the stronger large-scale velocity power in the shorter-reheating case leads to a larger GW amplitude in the infrared. This establishes a connection between the IR structure of the GW spectrum and the nonlinear redistribution of kinetic energy during the post-inflationary evolution.

For our benchmark parameters, the characteristic frequency and GW amplitude lie well beyond the sensitivity ranges of current and planned GW observatories such as LISA, DECIGO, and BBO. This, however, is not a generic limitation of the mechanism. The characteristic frequency is determined by the physical scale associated with the end of inflation and the subsequent reheating history, and therefore depends on both the inflationary energy scale and the reheating dynamics. In particular, inflationary scenarios with lower characteristic energy scales can shift the GW spectrum towards lower frequencies, potentially bringing the signal into the frequency bands accessible to future GW observatories. The GW spectrum can therefore, in suitable low-scale inflationary scenarios, provide a probe of the post-inflationary reheating dynamics.

\color{black}
\section{Conclusions}

In this paper, we report the first three-dimensional numerical simulations of magnetohydrodynamic turbulence during the post-inflationary reheating era up to our knowledge. The background dynamics are driven by a coherently oscillating inflaton field decaying into radiation. We find that the continuous buildup of the radiation bath leads to a nontrivial growth of the plasma kinetic energy. We have analysed how reheating alters the evolution of velocity, vorticity, and magnetic fields compared to previous MHD simulations, which assume an instantaneous reheating scenario~\cite {RoperPol:2023dzg, Kahniashvili:2015msa, Brandenburg:1996fc}. Furthermore, we present an analysis of the resulting gravitational waves, demonstrating the impact of the velocity perturbation and magnetic fields in the homogeneous reheating scenario.

We find that the continuous decay of the inflaton field acts as a steady source of energy to the radiation plasma. It maintains the turbulent motion up to the end of reheating. 
This energy from the inflaton decay is directly transferred to the homogeneous $k=0$ mode of the radiation density and does not directly source non-zero $k$ mode perturbations. Due to the relative density perturbation, the velocity fluctuation amplitude initially increases in the system. As a consequence of relative density perturbation and the Smagorinsky viscosity, the velocity field develops significantly richer filamentary structures and stronger small-scale fluctuations in the homogeneous reheating scenarios. This behaviour is quantitatively confirmed by the velocity and vorticity spectra, which show that energy at small scales does not decay significantly throughout the dynamical evolution of the fields.
%SD: add non-instantaneous case for density

Despite the substantial enhancement of kinetic energy, this enhancement does not translate into magnetic-field amplification in our simulations. In our current setup, we assume that the inflaton energy density is homogeneous. Therefore, the inflaton energy contributes to the background radiation energy density and does not amplify the density fluctuation very much. 
%\arp{how? should only contribute to the background}. 
As a result, in the homogeneous reheating approximation, inflaton energy density does not generate additional velocity gradients or small-scale inhomogeneities. Consequently, the magnetic-field evolution remains largely unchanged from that of the instantaneous reheating MHD scenario. 
%SD: add non-instantaneous case for velocity.

Our results establish that reheating is an active dynamical phase that significantly affects the evolution of subhorizon primordial plasma fluctuations, rather than serving merely as a transition between inflation and the radiation-dominated Universe. In this homogeneous reheating model, the inflaton does not directly contribute to producing inhomogeneities in the velocity field. Introducing a channel that directly converts the inflaton energy into the inhomogeneous radiation density and velocity perturbations may lead to magnetic-field amplification. Our current work serves as a starting point to investigate this process further in future work. More broadly, our study opens a new direction for understanding the interplay among turbulence, plasma dynamics, and magnetic field evolution in the early Universe. These effects may have important implications for magnetogenesis, gravitational-wave production, and other probes of non-instantaneous reheating histories.

\begin{acknowledgements}

SS would like to thank Dr Nishant Kumar Singh, Dr Ramkishor Sharma, Prof. Axel Brandenburg, and Chandranathan Anandavijayan for their useful feedback and comments on our work. SM gratefully acknowledges the financial support provided by the Council of Scientific and Industrial Research (CSIR), Ministry of Science and Technology, Government of India (GoI), through the Senior Research Fellowship (File No.\ 09/731(0192)/2021-EMR-I). A part of the computations presented in this work was performed at the University of Geneva using the Baobab HPC service. SS and ARP acknowledge the support provided by the Baobab HPC service. ARP acknowledges support by the Swiss National Science Foundation (SNSF Ambizione grant \href{https://data.snf.ch/grants/grant/208807}{182044}). SD gratefully acknowledges the OzSTAR national facility at Swinburne University of Technology, where a significant part of the numerical modelling was performed. The OzSTAR programme receives funding in part from the Astronomy National Collaborative Research Infrastructure Strategy (NCRIS) allocation provided by the Australian Government and from the Victorian Higher Education State Investment Fund (VHESIF) provided by the Victorian Government. SS, SM, and ARP acknowledge the support of the Bernoulli Center during the ``{\it Generation, evolution, and observations of cosmological magnetic fields}'' program where the discussions on the numerical implementation of this work took place. SS and ARP acknowledge the support from CERN to host the {\sc Pencil Code} school and user meeting, where part of the numerical work presented in this paper was discussed. The authors are grateful for the support provided by ICTS during the program ``{\em Hearing beyond the standard model with cosmic sources of Gravitational Waves.}'' SS acknowledges the support of Nordita during the Winter School “\textit{Cosmological Magnetic Fields: Generation, Observation, and Modeling}”, where part of this work was presented and discussed. SS and DM also acknowledge the PARAM Kamrupa supercomputing facility at the Indian Institute of Technology Guwahati (IIT Guwahati), India, supported under the National Supercomputing Mission (NSM), Government of India.

\end{acknowledgements}

% \vspace{2mm}

% {\bf Data availability}---

% \arp{We should prepare the datasets to be made public in Zenodo, but we can discuss this after putting it on arXiv (and prepare reading/postprocessing notebooks with CosmoGW). Mention Pencil Code here too}

%=================
% \newpage
\bibliographystyle{apsrev4-1}
\bibliography{references}

 \newpage
% \null
% \null
% \newpage

\appendix

\begin{widetext}

\begin{center}
{\large \textbf{Appendix}}
\end{center}

In this appendix, we present the detailed formulation of the two-fluid reheating model in the presence of electromagnetic fields, followed by the homogeneous reheating approach and a discussion of the initial conditions to support the main text.

%=====================================================
\section{Coupled Inflaton–Radiation–MHD System}
%=====================================================

%We describe reheating as a two-component system consisting of 
%a scalar inflaton field $\phi$ and a radiation fluid. 
%The total energy–momentum tensor is
%\begin{equation}
%T^{\mu\nu} = T^{\mu\nu}_\phi + T^{\mu\nu}_r .
%\end{equation}
%Total energy–momentum conservation,
%$\nabla_\mu T^{\mu\nu}=0$, allows for energy transfer between 
%the two components,
%\begin{equation}
%\nabla_\mu T^{\mu\nu}_\phi = -Q^\nu,
%\qquad
%\nabla_\mu T^{\mu\nu}_r = Q^\nu,
%\end{equation}
%where $Q^\nu$ parametrizes inflaton decay.

%%%%%%%%%%%%%%%%%%%%%%%%%%%%%%%%%%%%%%%%%%%%%%%%%%%%%%%%%%%%%%%%%%%%%%%%%%%%

To study the evolution of primordial magnetic fields and their possible influence on the background, it is imperative to derive the MHD equations. These equations provide a comprehensive framework describing the coupled dynamics of the plasma, electromagnetic fields, and background expansion. In this section, we derive the background MHD equations starting from the covariant conservation of the total energy-momentum tensor, which includes contributions from both the inflaton field and the radiation field coupled with the electromagnetic field.

We describe reheating as a two-component system consisting of a scalar inflaton field $\phi$ and a radiation fluid. The total energy–momentum tensor is
\begin{equation}
T^{\mu\nu} = T^{\mu\nu}_\phi + T^{\mu\nu}_r .
\end{equation}

It is well established that the dynamics of a minimally coupled scalar field can be described as a perfect fluid~\cite{PhysRevD.81.103511, RevModPhys.78.537, Faraoni:2012hn, Madsen:1988ph,
PhysRevD.2.2762,Brown:1992kc,Ballesteros:2016kdx,Faraoni:2022gry}.

Since the inflaton field is neutral and does not directly couple to the gauge field, the magnetic field interacts only with the radiation fields. In this scenario, we assume that significant magnetic fields are generated during inflation and that they couple to the radiation field, given that the inflaton is a scalar field. So the energy-momentum tensor of both fluids can be expressed as follows:
%The total energy-momentum tensor,  $T_\nu^\mu$, can be decomposed into two components: (i) the contribution from the scalar field, $T^{\mu(\phi)}_\nu$, and (ii) the radiation field, which also includes the electromagnetic (EM) field, denoted as $T^{\mu(\rm r)}_\nu$s are given by
\begin{subequations}
\begin{align}
%T^{\alpha\beta} &=(p+\rho)u^\alpha u^\beta + pg^{\alpha\beta} +\pi^{\alpha\beta} \\
T^{\mu \nu}_\phi &= (\rho_\phi + p_\phi)\, u^{\mu}_{\phi} u^{\nu}_{\phi} + p_\phi \delta^{\mu \nu}, \\
T^{\mu \nu}_r &= (\rho_{\rm r} + p_{\rm r} + \rho_{\rm em} + p_{\rm em})\, u^{\mu}_r u^{\nu}_r + (p_{\rm r} + p_{\rm em}) \delta^{\mu\nu}  - E^\mu E^\nu - B^\mu B^\nu.
\label{eq:stress_tensor}
\end{align} 
\end{subequations}
Where the electromagnetic field is treated as a perturbation, and the energy density and pressure of the electromagnetic field are denoted as $\rho_em$ and $p_{em}$. 
%Here $\rho_r$ is the energy density of the radiation fluid generated from the decay of the inflaton field, and $\pr$ is the pressure of the radiation field. 

We consider the perturbed metric in the Newtonian gauge, given by
\begin{align}
    ds^2 = a^2(\eta)\left[ -(1 + 2\Psi)d\eta^2 + (1 + 2\Phi)\delta_{ij}dx^i dx^j \right],
\end{align}
where $\Psi$ and $\Phi$ represent the scalar perturbations, and $a(\eta)$ is the scale factor.

Since the four-velocity $u^\mu$ must satisfy the normalisation condition $u^\mu u_\mu = -1$, the four-velocity components $u^\mu$ and $u_\mu$ in this background metric can be expressed as:
\begin{subequations}\label{eq:four_vel}
    \begin{align}
        u^\mu = a^{-1}\left[(1 - \Psi), u^i \right], \quad
        u_\mu = a\left[-(1 + \Psi), u^j \delta_{ij} \right],
    \end{align}
\end{subequations}
where $v^i$ represents the spatial velocity components in the co-moving frame.

\section{Perturbed Magnetohydrodynamics Equations}
%%%%%%%%%%%%%%%%%%%%%55
We begin with the conservation of the total energy-momentum tensor, given by
\begin{equation}
 \nabla_\mu T^{\mu\nu} = 0 ,
\end{equation}
which serves as the foundation for deriving the perturbed magnetohydrodynamic equations. While the total energy-momentum tensor is conserved, its individual components are not conserved independently due to the interactions among them. These interactions facilitate the exchange of energy and momentum between the inflaton and radiation fluids and can be expressed as follows:

\begin{equation}
\nabla_\mu T^{\mu\nu}_\phi = -Q^\nu, 
\qquad
\nabla_\mu T^{\mu\nu}_r = Q^\nu,
\end{equation}

where the four-vector \(Q^\nu\) characterises the transfer of energy and momentum from the inflaton field to the radiation component. The opposite signs in these equations ensure overall conservation, leading to

\begin{equation}
    \nabla_{\mu}T^{\mu\nu} = \sum_e \nabla_{\mu}T^{\mu\nu}_{e} = \sum_{e}Q^{\nu}_{e} = 0,
\end{equation}
where the index "e" stands for different interaction term.  In the perturbative reheating regime, the energy-momentum transfer can be written in terms of the inflaton decay rate $\Gamma_\phi$ and $\wphi$ denotes the equation of state of the inflaton as
\begin{equation}
 Q^\nu=\Gamma_{\phi}(1+\wphi)T_{\mu}^{\nu(\phi)}u^{\mu}_{\phi}
\end{equation}

%\subsection{Source Terms}

\subsection{Continuity Equations}

From the temporal component of the energy-momentum conservation for the inflaton, $\nabla_\mu T^{\mu 0}_\phi = -Q^0 $ we obtain the corresponding continuity equation for the inflaton energy density. At the perturbed level, it takes the form

\begin{equation}
\rho_\phi' + \partial_i q^i_\phi + 3\mathcal{H} (1 + \wphi) \rho_\phi 
+ \mathcal{H} \, q^i_\phi u_{i\phi}
= -a(1 + \Psi)(1 + \wphi)\Gamma_\phi \left( \rho_\phi - q^i_\phi u_{i \phi} \right)
\end{equation}

Where $q^i=(\rho+p)u^i$ is the momentum density, primes denote derivatives with respect to conformal time $\eta$, and $\mathcal{H} \equiv a'/a $ is the conformal Hubble parameter. Similarly, for the radiation component, the energy-momentum conservation leads to the perturbed continuity equation for radiation.

\begin{equation}
\rho_r' + \partial_i q^i_r + \rho_{\rm em}' + \partial_i (\rho_{\rm em} u^i_r ) + 4\mathcal{H} \rho_r + 2\mathcal{H} \rho_{\rm em} + \mathcal{H}(\rho_r + p_r + \rho_{\rm em}) u^i_r u_{i r} = a(1 + \Psi)(1 + \wphi)\Gamma_\phi \left( \rho_\phi - q^i_\phi u_{i \phi} \right).
\end{equation}

The last term in the source contribution on the right-hand side of the both equations is a higher-order contribution in the velocity perturbations and is therefore negligible at linear order in perturbations. Furthermore, assuming that anisotropic perturbations are subdominant, we set $ \Psi = 0 $. Under these approximations, the continuity equations for the inflaton and radiation reduce to,

\begin{align}
\rho_\phi' + \partial_i q^i_\phi + 3\mathcal{H} (1 + \wphi)\rho_\phi + \mathcal{H}\, q^i_\phi u_{i \phi}
&= -a(1 + \wphi)\Gamma_\phi \rho_\phi\, \\[6pt]
\rho_r' + \partial_i q^i_r + \rho_{\rm em}' + \partial_i(\rho_{\rm em} u^i_r) + 4\mathcal{H} \rho_r + 2\mathcal{H} \rho_{\rm em} 
+ \mathcal{H}(\rho_r + p_r + \rho_{\rm em}) u^i_r u_{i r}
&= a(1 + \wphi)\Gamma_\phi \rho_\phi\,
\end{align}

The electromagnetic energy density is defined as $ \rho_{\rm em} = \frac{1}{2} (E^2 + B^2)$. During the reheating epoch, the inflaton decays and rapidly transfers its energy to relativistic degrees of freedom, leading to the formation of a hot and dense plasma in the early Universe. This plasma exhibits very high electrical conductivity, which effectively suppresses the electric field component of the Universe. Consequently, the system approaches the ideal MHD limit, causing the electromagnetic energy density to be primarily dominated by the magnetic component. Therefore, the electromagnetic energy density can be approximated as $ \rho_{\rm em} \simeq \frac{1}{2} B^2 $.

Finally, in the co-moving frame, we obtain the following continuity equations for the inflaton and the radiation fluids in terms of logarithmic densities:

\begin{align}
     \frac{\partial \ln \trhop}{\partial \eta} 
 & = -(1 + \wphi) \left( \nabla\cdot \vup + \vup\cdot \nabla \ln \trhop \right)
 - a \Gp(1 + \wphi) \\[6pt]
    \frac{\partial \ln \trhor}{\partial \eta} 
 & = - \frac{4}{3}\left( \nabla\cdot \vur + \vur\cdot\nabla \ln \trhor \right) 
 + \frac{1}{\trhor}\left[\eta \tJ^2 + \vur \cdot (\tJ \times \tB)\right] + a^{2 - 3\wphi} \frac{\tilde{\rho}_\phi}{\tilde{\rho}_r} \Gp(1 + \wphi)
\end{align}

\subsection{Euler Equations}

Similarly, from the spatial component of the energy-momentum conservation, $\nabla_\mu T^{\mu i} = \pm Q^i$, we can calculate the velocity evolution equation for the inflaton and radiation fluids. In the co-moving frame, the Euler equations for the inflaton and the radiation fluids are given by,

\begin{align}
 \frac{\partial \vup}{\partial \eta} 
 & = - \vup \cdot \nabla \vup + \wphi \vup ( \nabla\cdot \vup + \vup \cdot\nabla \ln \trhop) - \frac{\wphi}{1+\wphi} \nabla \ln \trhop - \frac{(1-3\wphi)}{(1+3\wphi)}\mH \vup - a\Gp (1+\wphi) \vup
 \\[6pt]
 \frac{\partial \vur}{\partial \eta} 
 & = - \vur \cdot \nabla \vur - \frac{\vur}{3} ( \nabla\cdot \vur + \vur \cdot \nabla \ln \trhor ) - \frac{\vur}{\trhor} [\eta {\tJ}^2 + \vur (\tJ \times \tB)]\nn
 -\frac{1}{4} \nabla(\ln \trhor) + \frac{3}{4 \trhor} (\tJ\times \tB) \\
 &- \frac{\tilde{\rho}_{\phi}}{4 \Tilde{\rho}_r} a^{(2-3 \wphi)} \Gamma_{\phi}(1+\wphi) \vup
\end{align}

\subsection{Homogeneous Reheating}

In the homogeneous reheating scenario, we consider the inflaton condensate to be spatially uniform, so the inflaton velocity tends to zero. 
%In this scenario, the inflaton sector evolves solely through expansion and decay \textcolor{magenta}{of what?}. 
This represents the simplest configuration of the coupled inflaton-radiation system and is utilized to understand the effect of reheating in MHD evolution in the main text.

Physically, this indicates that the inflaton field does not exhibit any bulk motion. Instead, it gradually transfers its energy to the radiation fluid via the decay rate \(\Gamma_\phi\). As this transfer occurs, the radiation density increases, causing the plasma to become relativistic and conducting. In this conducting medium, the magnetic field evolves according to the MHD equations. In this limit, the inflaton continuity equation simplifies to:
\begin{align}
 \frac{\partial \ln \trhop}{\partial \eta} = - a \Gp(1 + \wphi).
\end{align}

This equation indicates that the inflaton energy density decreases steadily throughout the reheating process. The rate of this decrease is governed by the decay rate $\Gamma_\phi$ and the effective equation of state of the inflaton. Conversely, the radiation fluid receives energy from the inflaton and is also influenced by electromagnetic terms. The continuity equation of radiation plasma remains as follows:
\begin{align}
 \frac{\partial \ln \trhor}{\partial \eta} = - \frac{4}{3}\left( \nabla\cdot \vur + \vur\cdot\nabla \ln \trhor \right)  + \frac{1}{\trhor}\left[\eta \tJ^2 + \vur \cdot (\tJ \times \tB)\right]  
 + a^{2  - 3 \wphi} \frac{\tilde{\rho}_\phi}{\tilde{\rho}_r} \Gp(1 + \wphi).
\end{align}
The last term reflects the energy transferred from the inflaton to the radiation sector. it continuously feeds the plasma, allowing bulk motion to develop over time. As the inflaton velocity tends to zero, the radiation velocity obeys,
\begin{align}
 \frac{\partial \vur}{\partial \eta} 
 = - \vur \cdot \nabla \vur - \frac{\vur}{3} ( \nabla\cdot \vur + \vur \cdot \nabla \ln \trhor ) - \frac{\vur}{\trhor} [\eta {\tJ}^2 + \vur (\tJ \times \tB)] -\frac{1}{4} \nabla(\ln \trhor) + \frac{3}{4 \trhor} (\tJ\times \tB).
\end{align}

The homogeneous reheating limit is useful, as it isolates the role of the background source terms in the radiation energy density. Here, magnetic-field amplification is not driven by external shear but emerges from fluid motions generated self-consistently during reheating.

\begin{figure*}
    \centering
    \includegraphics[width=0.45\linewidth]{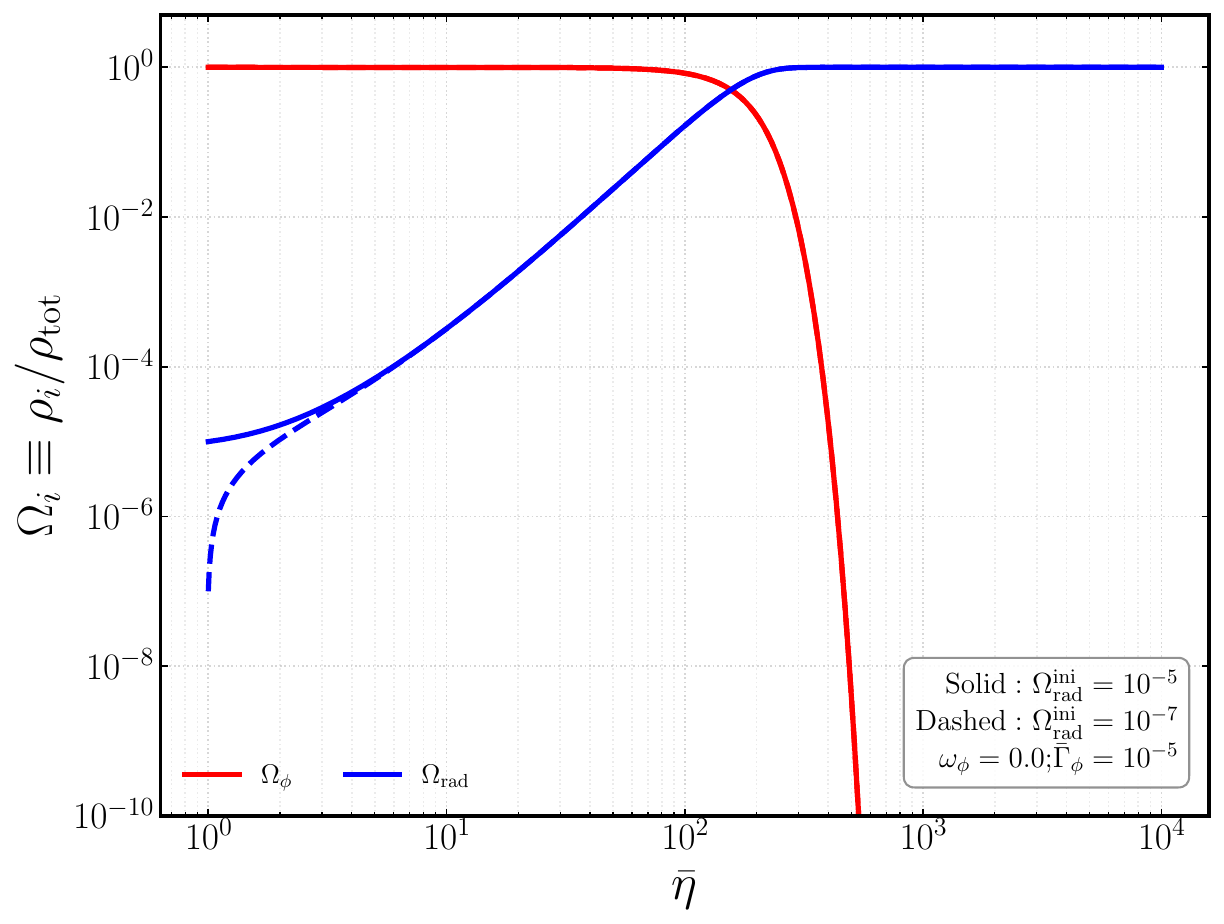}
     \includegraphics[width=0.45\linewidth]{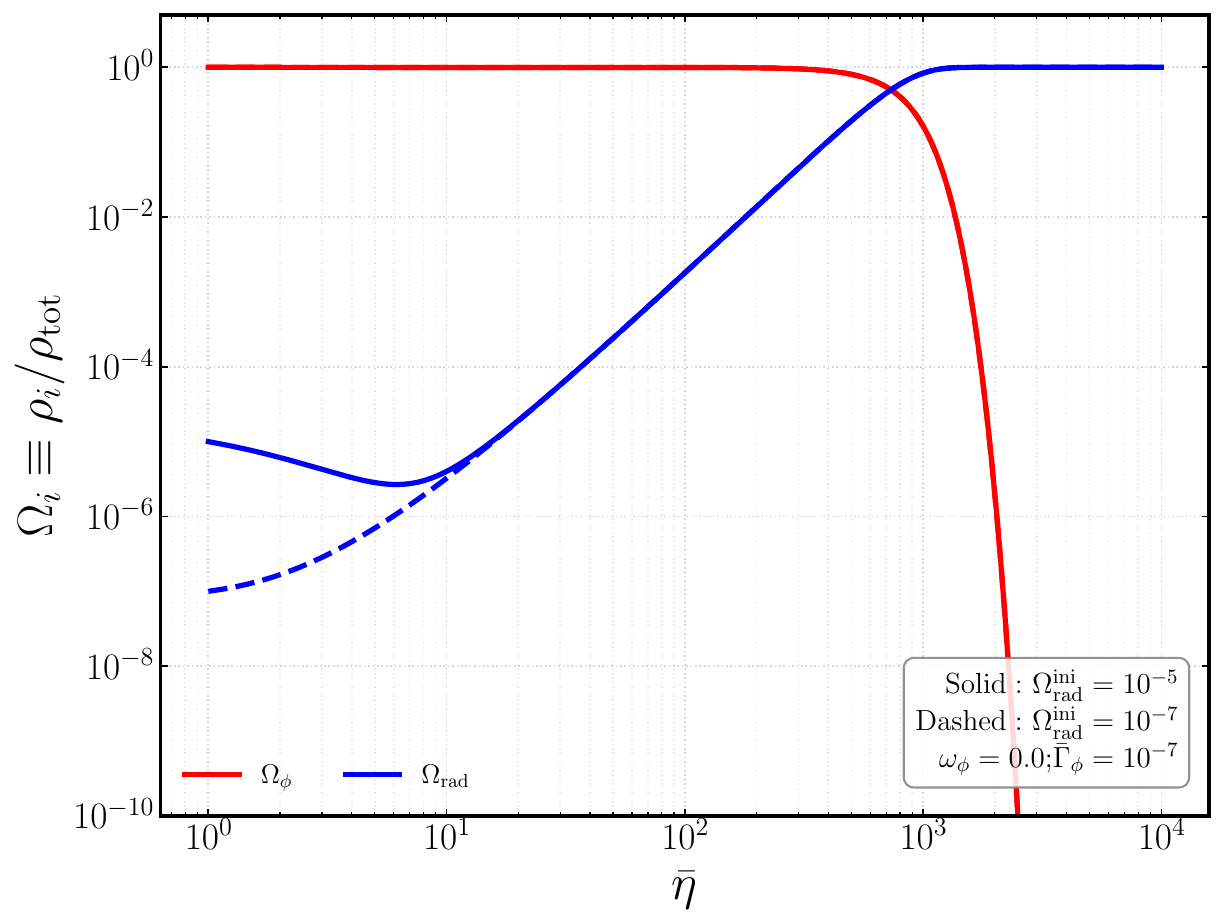}
    \caption{Evolution of the background energy densities of the inflaton and radiation fields as a function of normalised time for a matter-like reheating phase, $\wphi=0.0$, with $\tGp=10^{-5}$ and $\tGp=10^{-7}$.}
    \label{fig:initial_con}
\end{figure*}

To examine whether the duration of reheating depends on the initial radiation abundance, we plot the fractional energy densities of the inflaton and radiation fields, $\Ophi$ and $\Ora$, as functions of the normalised conformal time $\teta$ in Fig.~\ref{fig:initial_con}. The left and right panels correspond to $\tGp=10^{-5}$ and $\tGp=10^{-7}$, respectively, with the effective equation of state fixed at $\wphi=0.0$ throughout reheating. For each decay rate, we consider two initial radiation abundances: $\Ora^{\rm ini}=10^{-5}$, as adopted in our main simulations, shown by the solid curves, and $\Ora^{\rm ini}=10^{-7}$, shown by the dashed curves.

The comparison shows that, for the range of equations of state $\wphi\leq 1/3$ considered here, the initial radiation abundance has little effect on the duration of reheating. The main differences appear during the early evolution, when the relative importance of the pre-existing radiation and the radiation produced by inflaton decay depends on the initial conditions. In particular, when the initial radiation abundance is sufficiently small, the radiation produced by inflaton decay quickly dominates over the initial radiation component. The fractional radiation energy density then follows the scaling $\Ora\propto \tGp \teta^{\frac{5-3\wphi}{1+3\wphi}}$.

This behaviour is particularly evident in the right panel, where $\tGp=10^{-7}$. For $\Ora^{\rm ini}=10^{-5}$, the fractional radiation energy density decreases approximately as $\Ora\propto \teta^{-2}$ over $1\leq \teta\leq 10$. Since the inflaton decay rate is small, the production of radiation is initially inefficient, and the pre-existing radiation is diluted relative to the dominant inflaton component as the Universe expands. For $\wphi=0$, the inflaton energy density scales as $\rhop\propto a^{-3}$, whereas the radiation energy density scales as $\rhor\propto a^{-4}$. Consequently, the fractional radiation energy density decreases as $\Ora=\rhor/\rhoc\propto a^{-1}$ during this stage. By contrast, when $\Ora^{\rm ini}=10^{-7}$, the initial radiation abundance is sufficiently small that radiation produced by inflaton decay dominates the evolution from an early time, causing $\Ora$ to increase according to the scaling given above. Despite these differences at early times, the two cases converge to a similar reheating history, demonstrating that the duration of reheating is largely insensitive to the initial radiation abundance for the scenarios considered here.

\begin{figure*}
    \centering
    \includegraphics[width=0.48\linewidth]{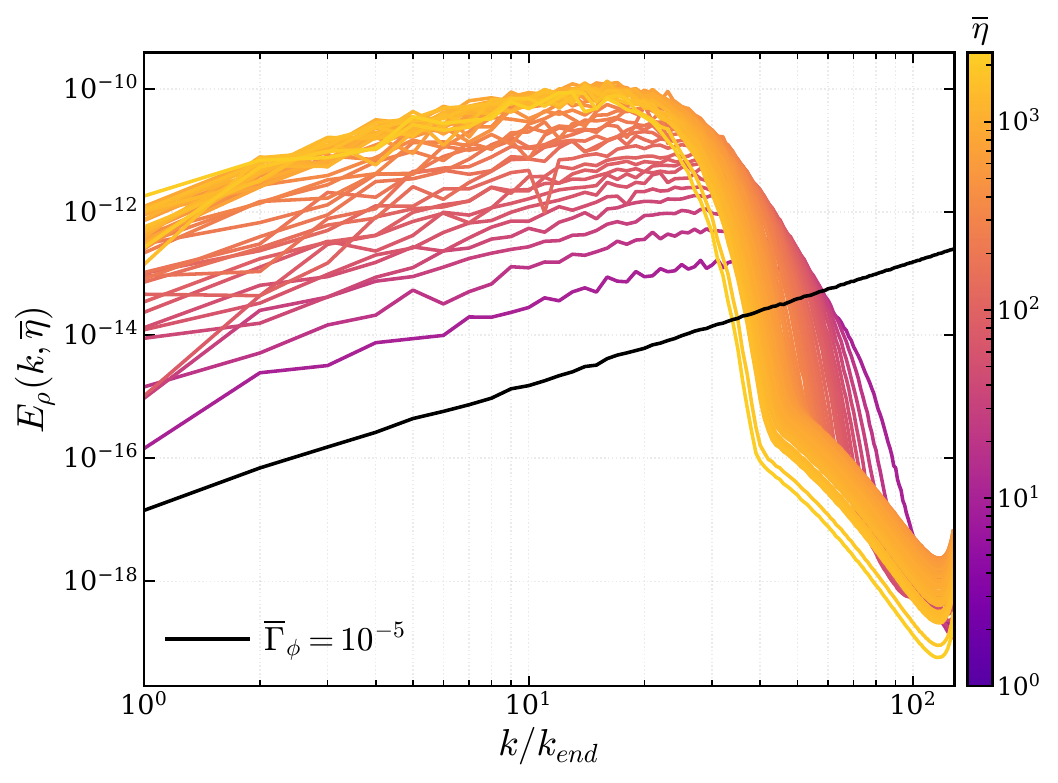}
    \includegraphics[width=0.48\linewidth]{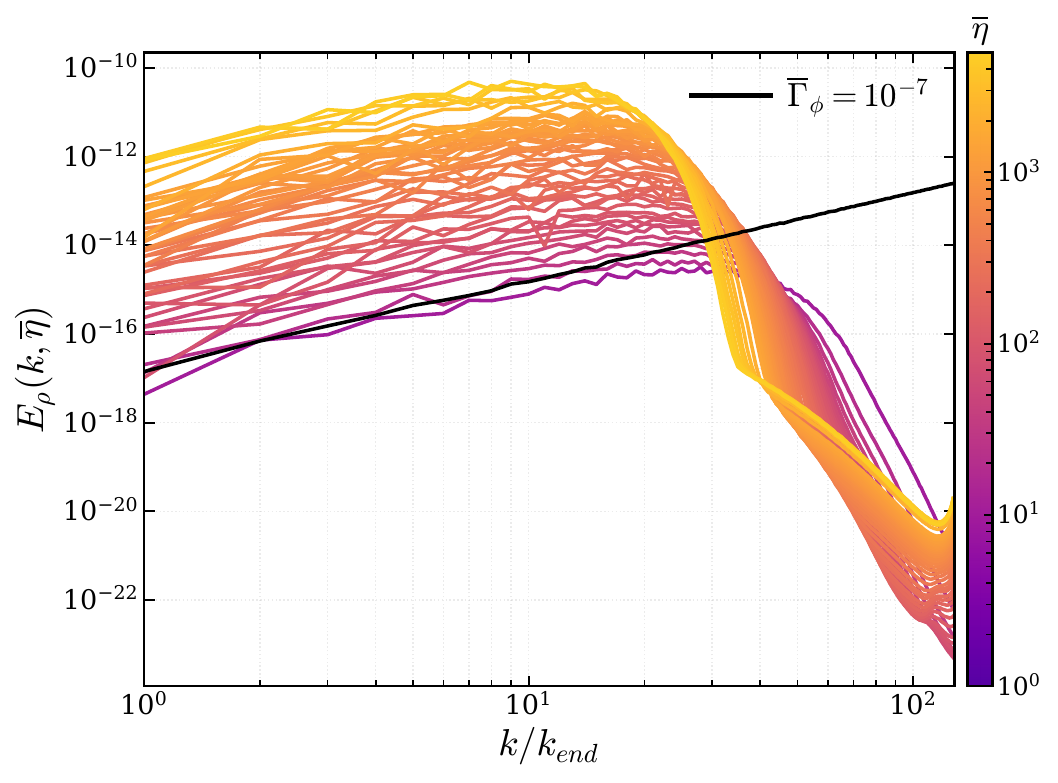}
    \caption{ Evolution of the radiation energy density spectrum for $ eos=1/3$. We have plotted the radiation energy density spectrum for $\bar \Gamma_\phi =10^{-5}$ in the left panel and the spectrum for $\bar \Gamma_\phi =10^{-7}$ in the right panel.}
    \label{fig:density_spectrum_eos0.3}
\end{figure*}

\begin{figure*}
    \centering
    \includegraphics[width=0.48\linewidth]{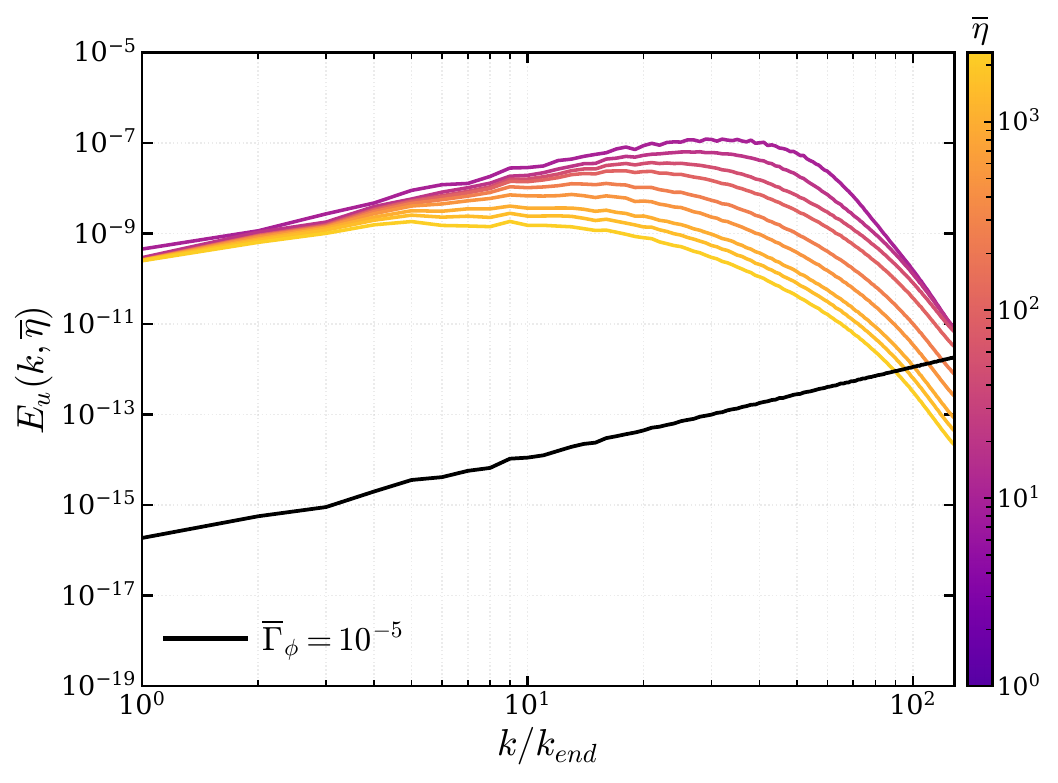}
    \includegraphics[width=0.48\linewidth]{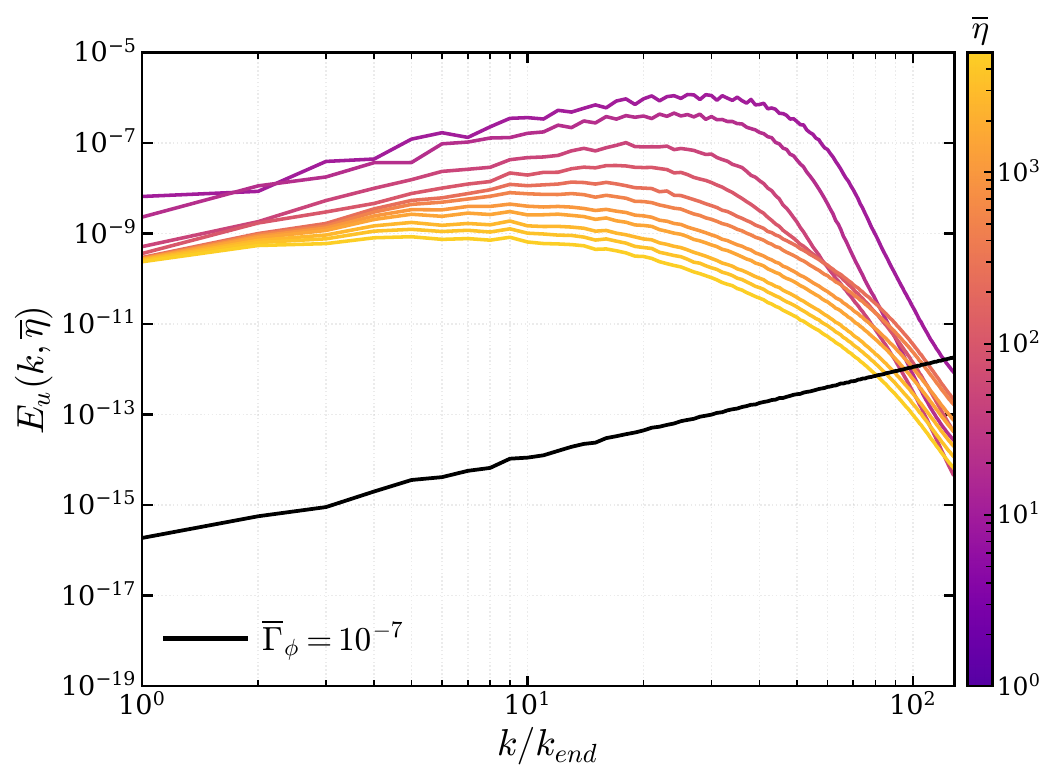}
    \caption{Evolution of the velocity spectrum of the radiation plasma, for $ eos=1/3$. In the left panel, we have plotted the velocity spectrum for $\bar \Gamma_\phi =10^{-5}$.  In the right panel, we have plotted the velocity spectrum for $\bar \Gamma_\phi =10^{-7}$.}
    %\purple{comparison of eos 0 and 1/3 plot}
    \label{fig:velocity_spectrum_eos0.3}
\end{figure*}

\subsection{Evolution  of density and velocity perturbation for equation of state $ 1/3$ }

We further examine the evolution of the radiation plasma for higher equations of state of the inflaton condensate. We have done a simulation considering $\wphi=1/3$. For $\wphi=0$, the density spectrum in the presence of reheating shows an overall increase in energy  in all scales. In contrast, for $\wphi=1/3$, the evolution of the density spectrum is more scale-dependent. In this case, the energy density increases at small wave numbers, while it decreases at larger wave numbers as the system evolves. The corresponding density spectra evolution for $\wphi=1/3$ are shown in Fig.~\ref {fig:density_spectrum_eos0.3}. Thus the $\wphi=1/3$ case exhibits a redistribution of radiation energy density towards larger spatial scales.

This difference originates from the different energy transfer associated with the two inflaton equations of state. In our homogeneous reheating setup, the radiation source is spatially homogeneous and therefore does not directly generate finite-wavenumber perturbations. However, the source term $\mathcal S_r$ depends on $\wphi$, so the evolution of the homogeneous radiation background differs between the two cases. The energy injection rate from the inflaton to the radiation plasma is different for the two equations of state. This modifies the evolution of the radiation energy density and velocity fields. For $\wphi=1/3$, the resulting velocity perturbation leads to a suppression of radiation energy density at high wavenumbers together with an enhancement at lower wavenumbers.
%, whereas for $\wphi=0$ the density power grows over all scales.

The velocity spectra exhibit the same trend with respect to the decay rate, the dependence being much stronger for $\wphi=0$ than for $\wphi=1/3$. For the case of $\wphi=0$, the slower reheating rate ($\bar{\Gamma}_\phi=10^{-7}$), develops substantially larger velocity perturbation at intermediate wavenumbers than in the case of $\bar{\Gamma}_\phi=10^{-5}$. For $\wphi=1/3$ the velocity spectra for the two decay rates remain similar in both amplitude and shape. The power is again concentrated around the intermediate wavenumbers and is suppressed at large k as shown in Fig.~\ref{fig:velocity_spectrum_eos0.3}.

In general, the radiation energy density spectra for $\wphi=1/3$ and $\wphi=0$ have some differences but the velocity fluctuation evolution is similar. These differences are mainly due to a redistribution of the radiation energy over different scales. The velocity spectra also show qualitatively similar behaviour in both cases. Consequently, within our homogeneous reheating framework, changing the inflaton equation of state from $\wphi=0$ to $\wphi=1/3$ does not lead to a significant qualitative change in the velocity spectra or the subsequent evolution of the magnetic field and gravitational waves.

\end{widetext}

%%%%%%%%%%%%%%%%%%%%%%%%%%%%%%%%%%%%%%%%%%%%%%%%%%%%%%%%%%%%%%%%%%%%%%%%%%%%%%%%%%%%%%%%%%%%%

\end{document}